# Hierarchical sparse Bayesian learning for structural health monitoring with incomplete modal data

Yong Huang and James L. Beck*

Division of Engineering and Applied Science, California Institute of Technology, Pasadena, CA 91125

**ABSTRACT**

For civil structures, structural damage due to severe loading events such as earthquakes, or due to long-term environmental degradation, usually occurs in localized areas of a structure. A new sparse Bayesian probabilistic framework for computing the probability of localized stiffness reductions induced by damage is presented that uses noisy incomplete modal data from before and after possible damage. This new approach employs system modal parameters of the structure as extra variables for Bayesian model updating with incomplete modal data. A specific hierarchical Bayesian model is constructed that promotes spatial sparseness in the inferred stiffness reductions in a way that is consistent with the Bayesian Ockham razor. To obtain the most plausible model of sparse stiffness reductions together with its uncertainty within a specified class of models, the method employs an optimization scheme that iterates among all uncertain parameters, including the hierarchical hyper-parameters. The approach has four important benefits: (1) it infers spatially-sparse stiffness changes based on the identified modal parameters; (2) the uncertainty in the inferred stiffness reductions is quantified; (3) no matching of model and experimental modes is needed, and (4) solving the nonlinear eigenvalue problem of a structural model is not required. The proposed method is applied to two previously-studied examples using simulated data: a ten-story shear-building and the three-dimensional braced-frame model from the Phase II Simulated Benchmark problem sponsored by the IASC-ASCE Task Group on Structural Health Monitoring. The results show that the occurrence of false-positive and false-negative damage detection is clearly reduced in the presence of modeling error (differences between the real structural behavior and the model of it). Furthermore, the identified most probable stiffness loss ratios are close to their actual values.

*Corresponding author. Fax: +1 626 578 0124

*E-mail address:* jimbeck@caltech.edu (James L. Beck).

## 1. Introduction

With the general goal of improving the safety and reducing life-cycle costs of critical civil infrastructure, structural health monitoring (SHM) has attracted increasing research interest in the structural engineering community over the last three decades [1-6]. The interest is in developing automated sensor-based systems for accurately detecting, locating and assessing earthquake-induced structural weakening (or damage from other severe loading events such as hurricanes, impacts or explosions, or from progressive structural deterioration at an early stage in its evolution). The most important objective of a damage identification algorithm is to reliably issue an alarm if damage has occurred. An alarm is generally issued if some damage features shift from their healthy state values, which usually is determined by a damage index obtained from an unknown state of the structure deviating from the healthy state beyond some threshold. Defining a proper threshold is the critical challenge to establish a timely and reliable damage alarm [8-9].

False indication of damage falls into two types [10]: (1) False-positive damage indication, which means that the algorithm indicates damage although no real damage is present; (2) False-negative damage indication, which means that the algorithm does not detect real damaged components, i.e., reports them as undamaged. False negative detection is usually more critical because undetected damaged elements may lead to severe consequences, even resulting in structural collapse. On the other hand, false positive detections can needlessly heighten concern about safety, and lead to costly visual inspections by engineers. Recently, some researchers have investigated how to compute a proper threshold value in a rigorous manner in order to alleviate false positive (false alarm) and false negative (missed alarm) detections [8, 11]. However, novel methods still need to be explored for better damage alarm performance.

Another challenge for structural damage detection is that existing methods often require measurement information at locations corresponding to every degree of freedom (DOF) of a model of the structure, whereas, in reality, sensors are typically installed at only a limited number of locations, so the spatial distribution of the structural motion is not known completely. Therefore, it is impossible to exactly describe the current state of the structure by the limited information available in practice, and we have a state of uncertainty that can be better described probabilistically. Rather than considering only a point estimate for the model parameters, Bayesian inference considers all possible values of the parameters and explicitly treats modeling uncertainty, including quantification of parametric uncertainty, by treating the problem within a framework of plausible inference in the presence of incomplete information (Beck 2010). Therefore, it provides a promising way to locate structural damage, which may occur away from the sensor locations or be hidden from sight. Furthermore, being able to quantify the uncertainties of the structural model parameters accurately and appropriately is essential for a robust prediction of future safety and reliability of the structure. The Bayesian framework has been used previously for damage detection and assessment [2-3, 12-18].

In this article, we explore recent developments in sparse Bayesian learning [19-22] and Bayesian compressive sensing [23-25] to perform sparse stiffness loss inference based on changes in the identified modal parameters from the sensor data. The physical basis for exploring sparseness in this inverse problem (i.e., inferring stiffness losses due to damage based on dynamic sensor data) is that the damage induced by an earthquake typically occurs at a limited number of locations (in the absence of collapse). This is important prior information that can be exploited.

We have proposed previously to use a sparse Bayesian learning approach to tackle this stiffness inversion problem in which a specific hierarchical Bayesian model is employed that induces sparseness in order to improve the accuracy and robustness of damage detection and assessment [26]. This approach employs an optimization scheme that iterates among all uncertain parameters to obtain the most plausible values of spatially-sparse stiffness reductions together with their uncertainty, based on the information in the experimentally identified modal parameters from the current unknown state and the original healthy state. In this paper, we improve the theoretical formulation and illustrate the ability of the proposed method to update a structural model during a calibration stage for an undamaged building (Example 1 taken from [18]), and to accurately infer damage in a monitoring stage (Example 2 taken from [27]), by applying the method to noisy incomplete modal data in both cases.

## 2. FORMULATION

### *2.1 Structural model class and target problem*

For a structure of interest, we take a class of linear structural models that has $d$ degrees of freedom, a known mass matrix $\mathbf{M}$ based on structural drawings and an uncertain stiffness matrix $\mathbf{K}$ that is represented as a linear combination of $(n+1)$ stiffness matrices $\mathbf{K}_{j,}\ j = 1, \dots n$, as follows:

$\mathbf{K}(\boldsymbol{\theta})=\mathbf{K}_0+\sum_{j=1}^{n}\theta_j\mathbf{K}_j$ (1)

where the nominal substructure stiffness matrices $\mathbf{K}_j \in \mathbb{R}^{d\times d}, j=1,\dots,n,$ represent the nominal contribution of the $j^{th}$ substructure of the structure to the overall stiffness matrix $\mathbf{K}$ from a structural model (e.g. based on the finite-element method), and $\boldsymbol{\theta}=[\theta_1,\dots,\theta_n]\in\mathbb{R}^n$ are corresponding stiffness scaling parameters that represent the structural model parameter vector to be updated by dynamic data. The reduction of any $\theta_j,\ j=1,\dots n,$ corresponds to damage in the $j^{th}$ substructure. Since structural damage induced by severe loading event, such as an earthquake, typically occurs at a limited number of locations in the absence of structural collapse, $\Delta\boldsymbol{\theta}=\boldsymbol{\theta}-\widehat{\boldsymbol{\theta}}_u$ can be considered as a sparse vector with relative few non-zero components, where $\boldsymbol{\theta}$ and $\boldsymbol{\theta}_u$ are the stiffness scaling parameters for current (possibly damaged) and undamaged states, and $\widehat{\boldsymbol{\theta}}_u$ is the MAP (maximum a posteriori) estimate of $\boldsymbol{\theta}_u$ determined from the calibration test data. We assume that a linear dynamic model with classical normal modes is adequate for damage detection purposes because we use low-amplitude vibration data recorded by the structural monitoring system just before and after an earthquake. Under this hypothesis, a damping matrix need not be explicitly modeled since it does not affect the model mode shapes.

Suppose that $q$ sets of measured vibration time histories are available from the structure and $m$ modes of the structural system have been identified for each set of time histories so that we have a vector of identified (MAP) system natural frequencies $\widehat{\boldsymbol{\omega}}^2=[\widehat{\omega}_{1,1}^2,\dots,\widehat{\omega}_{1,m}^2,\widehat{\omega}_{2,1}^2,\dots,\widehat{\omega}_{q,m}^2]^T\in\mathbb{R}^{qm\times 1}$ and mode shapes $\widehat{\boldsymbol{\psi}}=\left[\widehat{\boldsymbol{\psi}}_{1,1}^T,\dots,\widehat{\boldsymbol{\psi}}_{1,m}^T,\widehat{\boldsymbol{\psi}}_{2,1}^T,\dots,\widehat{\boldsymbol{\psi}}_{q,m}^T\right]^T\in\mathbb{R}^{qms\times 1}$, where $\widehat{\boldsymbol{\psi}}_{r,i}\in\mathbb{R}^s$ gives the identified components of the system mode shape of the $i^{th}$ mode ($i=1,\dots,m$) at the $s$ measured DOF from the $r^{th}$ data segment ($r=1,\dots,q$). These modal parameters are assumed to be directly estimated from dynamic data using an appropriate modal identification procedure, such as MODE-ID [15,28], which does not use a structural model and identifies the MAP values of the natural frequencies, equivalent viscous damping ratios and mode shape components at the observed degrees of freedom.

The target problem of interest is to use an appropriate sparse Bayesian learning technique for the inverse problem of inferring the pattern of stiffness loss $\Delta\boldsymbol{\theta}$ from the noisy incomplete modal data $\widehat{\boldsymbol{\omega}}^2$ and $\widehat{\boldsymbol{\psi}}$, and to then use $\Delta\boldsymbol{\theta}$ to decide whether to issue a damage alarm without having to set any stiffness loss thresholds. Ideally, we would like to treat each structural member as a substructure so that we can infer from the dynamic data which, if any, members have been damaged by the severe loading event. However, the information available from the structure's local network of sensors will generally be insufficient to support a member-level resolution of stiffness loss from damage, so larger substructures may be necessary in order to reduce the number of stiffness scaling parameters $\theta_j$ in Equation (1). A tradeoff is therefore required between the number of substructures (and hence the resolution of the damage locations) and the reliability of the probabilistically-inferred damage state. By inducing sparseness in the inferred stiffness loss through sparse Bayesian learning, we expect higher-resolution damage localization while still producing reliable damage assessment.

One complication in applying sparse Bayesian learning is that it requires a model that gives the predicted output as a linear function of the model parameters for which sparseness is to be enforced but, despite the linearity in Equation (1), the model for the modal parameters characterizing the modal data $\hat{\boldsymbol{\omega}}^2$ and $\hat{\boldsymbol{\psi}}$ is a nonlinear function of the structural stiffness scaling parameter vector $\boldsymbol{\theta}$. In the following formulation, rather than directly tackling this challenging nonlinear inverse problem, we apply an efficient iterative procedure that involves a series of coupled linear regression problems and so provides an appropriate form for the sparse Bayesian learning method.

### *2.2 Hierarchical Bayesian modeling*

#### *2.2.1 Priors for system modal parameters and structural stiffness scaling parameters*

To represent the actual modal parameters of the structure, we introduce *system* natural frequencies $\boldsymbol{\omega}^2 = [\omega_1^2, \ldots, \omega_m^2]^T \in \mathbb{R}^{m\times 1}$ and real *system* mode shapes $\boldsymbol{\phi} = [\boldsymbol{\phi}_1^T, \ldots, \boldsymbol{\phi}_m^T]^T \in \mathbb{R}^{dm\times 1}$ at the same $d$ degrees of freedom as the structural model [3]. We do not assume that the system natural frequencies $\boldsymbol{\omega}^2$ and system mode shapes $\boldsymbol{\phi}$ satisfy the eigenvalue problem corresponding to any structural model specified by the parameters $\boldsymbol{\theta}$ because there will always be modeling approximations, so:

$$\left(\mathbf{K}(\boldsymbol{\theta}) - \omega_i^2\mathbf{M}\right)\boldsymbol{\phi}_i = \mathbf{e}_i \tag{2}$$

where the uncertain eigenvalue equation errors $\mathbf{e}_i \in \mathbb{R}^d,\ i = 1, \ldots, m,$ are for the $i^{th}$ system mode and the structural model specified by $\boldsymbol{\theta}$. They are modeled probabilistically as independent and identically distributed Gaussian vectors with zero mean and covariance matrix $\beta^{-1}\mathbf{I}_d = \mathrm{diag}(\beta^{-1}, \ldots, \beta^{-1})$. This joint probability model for $\mathbf{e}_{1,} \ldots, \mathbf{e}_m$ maximizes Shannon's information entropy (i.e. it gives the largest uncertainty) for the equation errors subject to the moment constraints: $\mathbf{E}[(\mathbf{e}_i)_k] = 0, \mathbf{E}[(\mathbf{e}_i)_k^2] = \beta^{-1}, k = 1, \dots, d, i = 1, \dots, m$ [29]. Eq. (2) is then used to create the following prior PDF conditional on $\beta$:

$$p(\boldsymbol{\omega}^2, \boldsymbol{\phi}, \boldsymbol{\theta}|\beta) = c_0(2\pi\beta^{-1})^{-dm/2}\exp\left\{-\frac{\beta}{2}\sum_{i=1}^{m}\left\|\left(\mathbf{K}(\boldsymbol{\theta}) - \omega_i^2\,\mathbf{M}\right)\boldsymbol{\phi}_i\right\|^2\right\} \tag{3}$$

where $c_0$ is a normalizing constant and $\|\cdot\|$ denotes the Euclidean norm, so $\|\mathbf{x}\|^2 = \mathbf{x}^T\mathbf{x}$. Note that the equation-error precision parameter $\beta$ in (3) allows for the explicit control of how closely the system and model modal parameters agree. However, it is difficult to choose an appropriate value a priori for β and this motivate the introduction later of a hierarchical Bayesian prior (see Equation 8), where an optimal value of $\beta$ is learned from the data. Notice that as $\beta \to \infty$, the system modal parameters become tightly clustered around the modal parameters corresponding to the structural model specified by $\boldsymbol{\theta}$, which are given by Eq. (2) with all $\mathbf{e}_i = 0$, that is, $(\mathbf{K}(\boldsymbol{\theta}) - \boldsymbol{\omega}^2\mathbf{M})\boldsymbol{\phi} = 0$. Note also that if $\boldsymbol{\theta}$ is specified then these model modal parameters are always the most plausible values a priori of the system modal parameters.

From (1), we see that the exponent in (3) is a quadratic in $\boldsymbol{\theta}$ and so (3) can be analytically integrated with respect to $\boldsymbol{\theta}$ to get the marginal prior PDF for the system modal parameters $[\boldsymbol{\omega}^2, \boldsymbol{\phi}]$ as:

$$p(\boldsymbol{\omega}^2, \boldsymbol{\phi}|\beta)$$
$$= c_0(2\pi\beta^{-1})^{-(dm-n)/2}|\mathbf{H}^T\mathbf{H}|^{-1/2}\exp\left\{-\frac{\beta}{2}(\mathbf{b}^T\mathbf{b} - \mathbf{b}^T\mathbf{H}\,(\mathbf{H}^T\mathbf{H})^{-1}\mathbf{H}^T\mathbf{b})\right\} \tag{4}$$

where the matrix $\mathbf{H}$ and parameter vector $\mathbf{b}$ are defined by:

$$\mathbf{H} = \begin{bmatrix} \mathbf{K}_1\boldsymbol{\phi}_1 & \cdots & \mathbf{K}_n\boldsymbol{\phi}_1 \\ \vdots & \ddots & \vdots \\ \mathbf{K}_1\boldsymbol{\phi}_m & \cdots & \mathbf{K}_n\boldsymbol{\phi}_m \end{bmatrix}_{dm\times n} \tag{5}$$

$$\mathbf{b} = \begin{bmatrix} \left(\omega_1^2\,\mathbf{M} - \mathbf{K}_0\right)\boldsymbol{\phi}_1 \\ \vdots \\ (\omega_m^2\mathbf{M} - \mathbf{K}_0)\boldsymbol{\phi}_m \end{bmatrix}_{dm\times 1} \tag{6}$$

We can then deduce the prior PDF for $\boldsymbol{\theta}$ conditional on system modal parameters $[\boldsymbol{\omega}^2, \boldsymbol{\phi}]$ from (3) and (4):

$$p(\boldsymbol{\theta}|\boldsymbol{\omega}^2, \boldsymbol{\phi}, \beta) = p(\boldsymbol{\omega}^2, \boldsymbol{\phi}, \boldsymbol{\theta}|\beta)/p(\boldsymbol{\omega}^2, \boldsymbol{\phi}|\beta)$$
$$= (2\pi\beta^{-1})^{-n/2}|\mathbf{H}^T\mathbf{H}|^{1/2}\exp\left\{-\frac{\beta}{2}(\boldsymbol{\theta} - (\mathbf{H}^T\mathbf{H})^{-1}\mathbf{H}^T\mathbf{b})^T\mathbf{H}^T\mathbf{H}(\boldsymbol{\theta} - (\mathbf{H}^T\mathbf{H})^{-1}\mathbf{H}^T\mathbf{b})\right\} \tag{7}$$
$$= \mathcal{N}(\boldsymbol{\theta}|(\mathbf{H}^T\mathbf{H})^{-1}\mathbf{H}^T\mathbf{b}, (\beta\mathbf{H}^T\mathbf{H})^{-1})$$

It remains to define a hyper-prior for hyperparameter $\beta$. We take a Gamma conjugate hyper-prior on β:

$$p(\beta|a_0, b_0) = \mathrm{Gam}(\beta|a_0, b_0) = \frac{{b_0}^{a_0}}{\Gamma(a_0)}\beta^{a_0-1}\exp(-b_0\beta) \tag{8}$$

*Remark 2.1:* The choice of prior PDF in (3) builds on an idea by Yuen et al. [18]. They chose the prior PDF $p(\boldsymbol{\omega}^2, \boldsymbol{\phi}|\boldsymbol{\theta}, \beta) = c_1 \exp\left\{-\frac{\beta}{2}\sum_{i=1}^{m}\left\|\left(\mathbf{K}(\boldsymbol{\theta}) - \omega_i^2\ \mathbf{M}\right)\boldsymbol{\phi}_i\right\|^2\right\}$ with $c_1$ as a normalizing constant. However, this PDF is not a normalized PDF over the modal parameter space unless $c_1$ is a function of the structural model parameters $\boldsymbol{\theta}$ and the equation-error precision $\beta$. Therefore, we do not have the equivalent of a linear regression equation for the parameter vector $\boldsymbol{\theta}$ and so Bayesian inference, including the sparse Bayesian learning scheme, is analytically intractable for $\boldsymbol{\theta}$. In order to provide an appropriate form for sparse Bayesian learning, we introduce a new Bayesian model with the prior PDF in (3) where $c_0$, unlike $c_1$, is a constant with respect to $\boldsymbol{\omega}^2, \boldsymbol{\phi}$ and $\boldsymbol{\theta}$, and a corresponding likelihood function of $\boldsymbol{\theta}$ is introduced later in (12) of Section 2.2.3.

*2.2.2 Likelihood functions for system modal parameters*

The MAP estimates from modal identification are taken as the "measured" natural frequencies $\hat{\boldsymbol{\omega}}^2 = [\hat{\omega}_{1,1}^2, \dots, \hat{\omega}_{1,m}^2, \hat{\omega}_{2,1}^2, \dots, \hat{\omega}_{q,m}^2]^T$ and mode shapes $\hat{\boldsymbol{\psi}} = \left[\hat{\boldsymbol{\psi}}_{1,1}^T, \dots, \hat{\boldsymbol{\psi}}_{1,m}^T, \hat{\boldsymbol{\psi}}_{2,1}^T, \dots, \hat{\boldsymbol{\psi}}_{q,m}^T\right]^T$ [15, 30]. The combination of prediction errors and measurement errors for the system modal parameters are modeled as zero-mean Gaussian variables with unknown variances. This maximum entropy probability model gives the largest uncertainty for these errors subject to the first two moment constraints [29]. Based on this Gaussian model, one gets a Gaussian likelihood function for the system modal parameters $\boldsymbol{\omega}^2$ and $\boldsymbol{\phi}$ based on the measured quantities $\hat{\boldsymbol{\omega}}^2$ and $\hat{\boldsymbol{\psi}}$:

$$p\left(\hat{\boldsymbol{\omega}}^2, \hat{\boldsymbol{\psi}}|\boldsymbol{\omega}^2, \boldsymbol{\phi}, \boldsymbol{\theta}\right) = p(\hat{\boldsymbol{\omega}}^2|\boldsymbol{\omega}^2)p\left(\hat{\boldsymbol{\psi}}|\boldsymbol{\phi}\right)$$

$$= \mathcal{N}(\hat{\boldsymbol{\omega}}^2|\mathbf{T}\boldsymbol{\omega}^2, \mathbf{E})\ \mathcal{N}\left(\hat{\boldsymbol{\psi}}|\boldsymbol{\Gamma}\boldsymbol{\phi}, \eta^{-1}\mathbf{I}_{qms}\right) \tag{9}$$

where the selection matrix $\boldsymbol{\Gamma} \in \mathbb{R}^{qms \times dm}$ with "1s" and "0s" picks the observed degrees of freedom of the whole "measured" mode shape data set from the system mode shapes; $\mathbf{T} = \left[\mathbf{I}_{m,} \dots, \mathbf{I}_m\right]^T \in \mathbb{R}^{qm \times m}$ is the transformation matrix between the vector of $q$ sets of identified natural frequencies $\hat{\boldsymbol{\omega}}^2$ and the system natural frequencies $\boldsymbol{\omega}^2$; $\mathbf{E} = \text{block diag}\left(\mathbf{E}_1, \dots, \mathbf{E}_q\right)$ is a block diagonal matrix with the diagonal block matrices $\mathbf{E}_r = \text{diag}(\rho_1^{-1}, \dots, \rho_m^{-1}), r = 1, \dots, q$; $\mathbf{I}_m$ and $\mathbf{I}_{qms}$ denote the identity matrices of corresponding size; and $\boldsymbol{\rho} = [\rho_1, \dots, \rho_m]^T$ and $\eta$ are prescribed precision parameters for the predictions of the identified natural frequencies $\hat{\boldsymbol{\omega}}^2$ and mode shapes $\hat{\boldsymbol{\psi}}$ from the system modal parameters. In a hierarchical manner, exponential priors are placed on the parameters $\rho_i$ and $\eta$:

$$p(\rho_i|\tau_i) = \text{Exp}(\rho_i|\tau_i) = \tau_i \exp(-\tau_i\rho_i),\ i = 1, \dots, m; \qquad p(\eta|\nu) = \text{Exp}(\rho|\nu) = \nu \exp(-\nu\eta) \tag{10}$$

which are the maximum entropy priors with support $[0, \infty)$ for given mean values $\tau_i^{-1}$ and $\nu^{-1}$ of $\rho_i$ and $\eta$, respectively. Then the prior PDF for the parameter vector $\boldsymbol{\rho}$ is given to:

$$p(\boldsymbol{\rho}|\boldsymbol{\tau}) = \prod_{i=1}^{m} p(\rho_i|\tau_i) = \prod_{i=1}^{m} \tau_i \cdot \exp(-\sum_{i=1}^{m} \tau_i\rho_i) \tag{11}$$

where $\boldsymbol{\tau} = [\tau_1, \dots, \tau_m]^T$.

*2.2.3 Likelihood function for structural stiffness scaling parameters*

During the calibration stage, we use Bayesian updating of the structural model based on the identified modal parameters from tests of the undamaged structure to find the MAP structural stiffness scaling parameter $\hat{\boldsymbol{\theta}}_u$ and its corresponding uncertainty. For this stage, we assume that there is a unique MAP estimate $\hat{\boldsymbol{\theta}}_u$ due to the large amount of time-domain vibration data and identified modal parameters that can be collected because there is no rush. During the monitoring stage, we choose the MAP value $\hat{\boldsymbol{\theta}}_u$ from the calibration stage as pseudo-data for $\boldsymbol{\theta}$ and define a likelihood function to exploit the information that any stiffness changes $\Delta\boldsymbol{\theta} = \boldsymbol{\theta} - \hat{\boldsymbol{\theta}}_u$ should be a sparse vector (most of its components zero) for a structure in the absence of collapse. This is accomplished by incorporating the automatic relevance determination (ARD) concept [19,22] for $\Delta\boldsymbol{\theta}$: $p(\Delta\boldsymbol{\theta}|\boldsymbol{\alpha}) = \mathcal{N}(\Delta\boldsymbol{\theta}|\mathbf{0}, \mathbf{A})$ with $\mathbf{A} = \text{diag}(\alpha_1, \dots, \alpha_n)$, where each of the hyper-parameters $\alpha_j$ is the prediction-error variance for $\Delta\theta_j$. This choice is motivated by the closely-related sparse Bayesian learning framework which is known to provide an effective tool for pruning large numbers of irrelevant or redundant features in a linear regression model that are not supported by the data [19,26]. In sparse Bayesian learning, the ARD concept is used in the prior but here we use it in the likelihood function for $\boldsymbol{\theta}$, along with the prior on $\boldsymbol{\theta}$ in (7). This choice still leads to a sparse representation of the parameter change vector $\Delta\boldsymbol{\theta}$ during the optimization of the hyper-parameter vector $\boldsymbol{\alpha}$ using the evidence maximization strategy of [31].

The likelihood function for $\boldsymbol{\theta}$ based on the ARD concept is expressed as:

$$p(\hat{\boldsymbol{\theta}}_u|\boldsymbol{\theta}, \boldsymbol{\alpha}) = \mathcal{N}(\hat{\boldsymbol{\theta}}_u|\boldsymbol{\theta}, \mathbf{A}) = \prod_{j=1}^{n} \mathcal{N}(\hat{\theta}_{u,j}|\theta_j, \alpha_j) \tag{12}$$

To promote sparseness of the parameter vector $\Delta\boldsymbol{\theta}$ even more strongly, we take the following exponential hyper-prior PDF for $\boldsymbol{\alpha}$:

$$p(\boldsymbol{\alpha}|\lambda) = \prod_j^n p(\alpha_j|\lambda) = \prod_j^n \lambda \exp(-\lambda\alpha_j) = \lambda^n \exp\left(-\lambda \sum_j^n \alpha_j\right) \tag{13}$$

Finally, we model the uncertainty in $\lambda$ by an exponential hyper-prior:

$$p(\lambda|\zeta) = \mathrm{Exp}(\lambda|\zeta) = \zeta\exp(-\zeta\lambda) \quad (14)$$

Thus, a hierarchical Bayesian prior is defined for the structural stiffness scaling parameter vector $\boldsymbol{\theta}$.

<u>*Remark 2.2:*</u> For Bayesian sparsity modelling, the Laplace distribution $p(\widehat{\boldsymbol{\theta}}_u|\boldsymbol{\theta},\lambda) = \int p(\widehat{\boldsymbol{\theta}}_u|\boldsymbol{\theta},\boldsymbol{\alpha})\, p(\boldsymbol{\alpha}|\lambda) d\boldsymbol{\alpha}$ $= \frac{\lambda^{n/2}}{2^n}\exp\left(-\sqrt{\lambda}\left\|\widehat{\boldsymbol{\theta}}_u - \boldsymbol{\theta}\right\|_1\right)$ from (11) and (12) is desirable since it leads to a Bayesian MAP estimation that is equivalent to the $l_1$ regularization formulation that strongly enforces sparseness [24,32-34]. However, this Laplace likelihood is not conjugate to the Gaussian prior PDF in (7) and so the posterior PDF cannot be determined analytically from Bayes' theorem. The hierarchical Bayesian modelling of (12-14) is used instead; the first two stages (12-13) play a similar role to the Laplace likelihood (but without integrating over $\boldsymbol{\alpha}$) and the last stage in (14) is embedded to penalize values of $\lambda$ that are too large and so avoid $\boldsymbol{\theta}$ parameter vectors that are too sparse. In this hierarchical formulation, we note that the exponential hyper-prior in (13) is for the variance $\boldsymbol{\alpha}$ for sparseness promotion and not for the precision $\boldsymbol{\alpha}^{-1}$ that is usually used in sparse Bayesian learning [19].

*2.2.4 Joint posterior PDF for hierarchical Bayesian model*

By combining all the stages of the hierarchical Bayesian model, the joint posterior PDF of all the uncertain-valued parameters conditional on the observed quantities can be constructed. From Bayes' theorem, it is expressed as:

$$p(\boldsymbol{\omega}^2,\boldsymbol{\rho},\tau,\boldsymbol{\phi},\eta,\nu,\boldsymbol{\theta},\boldsymbol{\alpha},\lambda,\zeta,\beta|\widehat{\boldsymbol{\omega}}^2,\widehat{\boldsymbol{\psi}},\widehat{\boldsymbol{\theta}}_u,a_0,b_0)$$
$$\propto p(\widehat{\boldsymbol{\omega}}^2|\boldsymbol{\omega}^2,\boldsymbol{\rho})p(\widehat{\boldsymbol{\psi}}|\boldsymbol{\phi},\eta)p(\widehat{\boldsymbol{\theta}}_u|\boldsymbol{\theta},\boldsymbol{\alpha})p(\boldsymbol{\omega}^2,\boldsymbol{\phi},\boldsymbol{\theta}|\beta)p(\boldsymbol{\rho}|\boldsymbol{\tau})p(\eta|\nu)p(\boldsymbol{\alpha}|\lambda)p(\lambda|\zeta)p(\beta|a_0,b_0) \quad (15)$$

where the PDF is constructed by combining the different levels of the hierarchical Bayesian model. Note that we have omitted the product of PDFs $p(\boldsymbol{\tau})p(\nu)p(\zeta)$ for notational convenience, since they are all chosen as broad uniform priors and so are constant.

Hierarchical Bayesian models make use of the property of the conditional dependencies in the joint probability model and a graphical model representation is demonstrated in Figure 1, where each arrow denotes the generative model (conditional dependencies). Note that the key idea of the formulation is demonstrated in the first five blocks from the left, i.e., the pseudo evidence for the pseudo data $\widehat{\boldsymbol{\theta}}_u$ is maximized with respect to the hyper-parameters $\boldsymbol{\alpha}$, $\lambda$ and $\zeta$, which forces many of the $\alpha_j, j = 1, \dots, n,$ to approach zero during the optimization and the corresponding $\theta_j$'s become equal to their value for the undamaged state. This forces the inferred stiffness reductions to be spatially sparse in a way that is consistent with the Bayesian Ockham razor [31,35]. A stiffness scaling parameter $\theta_j$ is changed from its calibration value $\hat{\theta}_{u,j}$ only if the posterior probability of the model class with $\theta_j$ fixed at $\hat{\theta}_{u,j}$ is less than a model class with $\theta_j$ free to be updated.

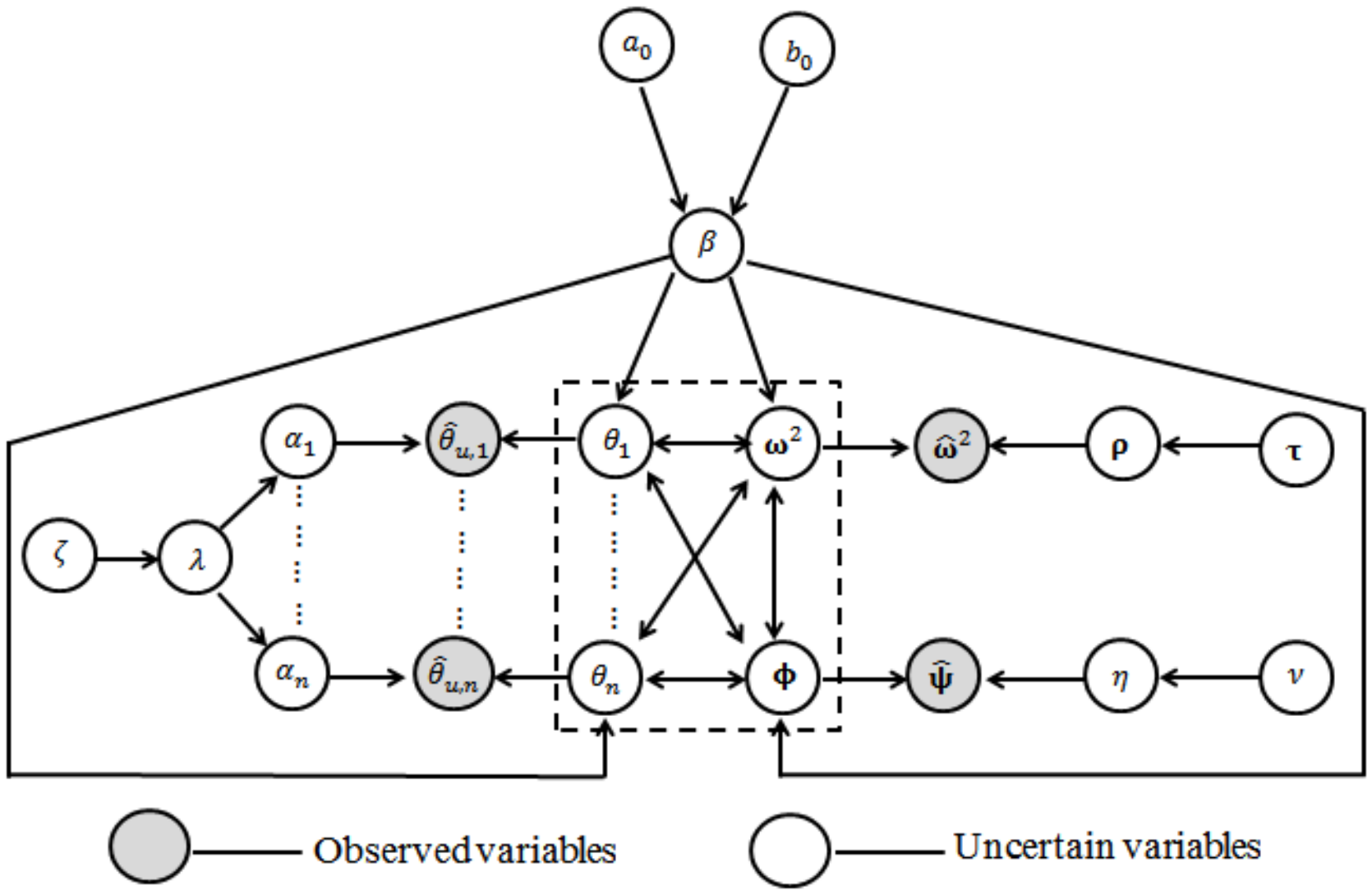

**Fig. 1.** The directed acyclic graph of the hierarchical Bayesian model.

### *2.3 Bayesian inference*

#### *2.3.1 Approximation of the full posterior PDF using Laplace's method*

The structural stiffness scaling parameter $\boldsymbol{\theta}$ is of key interest for damage detection, along with its associated hyper-parameters $\boldsymbol{\delta} = [\boldsymbol{\alpha}^T, \lambda, \zeta]^T$ as shown in (12-14). For convenience, we denote the other uncertain parameters as $\boldsymbol{\xi} = [\beta, (\boldsymbol{\omega}^2)^T, \boldsymbol{\rho}^T, \boldsymbol{\tau}^T, \boldsymbol{\phi}^T, \eta, \nu]^T$. The posterior PDF over all uncertain parameters is then expressed as

$$p(\boldsymbol{\xi}, \boldsymbol{\theta}, \boldsymbol{\delta} | \hat{\boldsymbol{\omega}}^2, \hat{\boldsymbol{\psi}}, \hat{\boldsymbol{\theta}}_u, a_0, b_0) = \frac{p(\boldsymbol{\xi}, \boldsymbol{\theta}, \boldsymbol{\delta}, \hat{\boldsymbol{\omega}}^2, \hat{\boldsymbol{\psi}}, \hat{\boldsymbol{\theta}}_u | a_0, b_0)}{p(\hat{\boldsymbol{\omega}}^2, \hat{\boldsymbol{\psi}}, \hat{\boldsymbol{\theta}}_u | a_0, b_0)} \tag{16}$$

However, this posterior PDF is nearly always intractable, since the denominator $(\hat{\boldsymbol{\omega}}^2, \hat{\boldsymbol{\psi}}, \hat{\boldsymbol{\theta}}_u | a_0, b_0)$ in (16) is given by an integral that cannot be computed analytically:

$$(\hat{\boldsymbol{\omega}}^2, \hat{\boldsymbol{\psi}}, \hat{\boldsymbol{\theta}}_u | a_0, b_0) = \int p(\boldsymbol{\xi}, \boldsymbol{\theta}, \boldsymbol{\delta}, \hat{\boldsymbol{\omega}}^2, \hat{\boldsymbol{\psi}}, \hat{\boldsymbol{\theta}}_u | a_0, b_0) d\boldsymbol{\xi} d\boldsymbol{\theta} d\boldsymbol{\delta} \tag{17}$$

Nevertheless, a hierarchical Bayesian procedure combined with Laplace's asymptotic approximation can provide an effective approximation of the full posterior PDF.

We treat $\boldsymbol{\delta}$ as a 'nuisance' parameter vector and attempt to integrate it out to get the posterior for $[\boldsymbol{\xi}, \boldsymbol{\theta}]$:

$$p(\boldsymbol{\xi}, \boldsymbol{\theta}|\hat{\boldsymbol{\omega}}^2, \hat{\boldsymbol{\psi}}, \hat{\boldsymbol{\theta}}_u) = \int p(\boldsymbol{\xi}, \boldsymbol{\theta}, \boldsymbol{\delta}|\hat{\boldsymbol{\omega}}^2, \hat{\boldsymbol{\psi}}, \hat{\boldsymbol{\theta}}_u) d\boldsymbol{\delta}$$

$$= \int p(\boldsymbol{\xi}, \boldsymbol{\theta}|\boldsymbol{\delta}, \hat{\boldsymbol{\omega}}^2, \hat{\boldsymbol{\psi}}, \hat{\boldsymbol{\theta}}_u) p(\boldsymbol{\delta}|\hat{\boldsymbol{\omega}}^2, \hat{\boldsymbol{\psi}}, \hat{\boldsymbol{\theta}}_u)\, d\boldsymbol{\delta} \tag{18}$$

where, from now on, we leave the conditioning on $(a_0, b_0)$ implicit in the PDFs. Using Bayes' theorem, we get a modified form of (15):

$$p(\boldsymbol{\xi}, \boldsymbol{\theta}|\boldsymbol{\delta}, \hat{\boldsymbol{\omega}}^2, \hat{\boldsymbol{\psi}}, \hat{\boldsymbol{\theta}}_u) \propto p(\hat{\boldsymbol{\omega}}^2, \hat{\boldsymbol{\psi}}, \hat{\boldsymbol{\theta}}_u|\boldsymbol{\delta}, \boldsymbol{\xi}, \boldsymbol{\theta}) p(\boldsymbol{\xi}, \boldsymbol{\theta}|\boldsymbol{\delta})$$

$$= p(\hat{\boldsymbol{\omega}}^2|\boldsymbol{\omega}^2, \rho) p(\hat{\boldsymbol{\psi}}|\boldsymbol{\phi}, \eta) p(\hat{\boldsymbol{\theta}}_u|\boldsymbol{\theta}, \boldsymbol{\alpha})\, p(\boldsymbol{\omega}^2, \boldsymbol{\phi}, \boldsymbol{\theta}|\beta) p(\rho|\tau) p(\eta|\nu) p(\beta|a_0, b_0) \tag{19}$$

Assuming that the posterior $p(\boldsymbol{\delta}|\hat{\boldsymbol{\omega}}^2, \hat{\boldsymbol{\psi}}, \hat{\boldsymbol{\theta}}_u)$ in (18) has a unique maximum at $\tilde{\boldsymbol{\delta}}$ (the MAP value of $\boldsymbol{\delta}$), we apply Laplace's asymptotic approximation [14] to (18) to get:

$$p(\boldsymbol{\xi}, \boldsymbol{\theta}|\hat{\boldsymbol{\omega}}^2, \hat{\boldsymbol{\psi}}, \hat{\boldsymbol{\theta}}_u) \approx p(\boldsymbol{\xi}, \boldsymbol{\theta}|\tilde{\boldsymbol{\delta}}, \hat{\boldsymbol{\omega}}^2, \hat{\boldsymbol{\psi}}, \hat{\boldsymbol{\theta}}_u) \tag{20}$$

where $\tilde{\boldsymbol{\delta}} = \arg\max p(\boldsymbol{\delta}|\hat{\boldsymbol{\omega}}^2, \hat{\boldsymbol{\psi}}, \hat{\boldsymbol{\theta}}_u)$.

*2.3.2 Bayesian inference for the posterior PDF of* $p(\boldsymbol{\xi}, \boldsymbol{\theta}|\tilde{\boldsymbol{\delta}}, \hat{\boldsymbol{\omega}}^2, \hat{\boldsymbol{\psi}}, \hat{\boldsymbol{\theta}}_u)$

*MAP estimation by iterative optimizations*

The optimal (MAP) values $[\tilde{\xi}, \tilde{\theta}]$ of the unknown model parameters $[\boldsymbol{\xi}, \boldsymbol{\theta}]$ can be found by minimizing the negative logarithm of the posterior PDF $p(\boldsymbol{\xi}, \boldsymbol{\theta}|\tilde{\boldsymbol{\delta}}, \hat{\boldsymbol{\omega}}^2, \hat{\boldsymbol{\psi}}, \hat{\boldsymbol{\theta}}_u)$ given by (19) with $\boldsymbol{\delta}$ fixed at its MAP value $\tilde{\boldsymbol{\delta}}$, so that $\tilde{\mathbf{A}} = \mathrm{diag}(\tilde{\alpha}_1, \dots, \tilde{\alpha}_n)$ is fixed. This leads to minimization of:

$$J(\boldsymbol{\xi}, \boldsymbol{\theta}) = (1 - a_0)\log\beta + b_0\beta - \frac{q}{2}\sum_{i=1}^{m} \log\rho_i + \frac{1}{2}(\hat{\boldsymbol{\omega}}^2 - \mathbf{T}\boldsymbol{\omega}^2)^T \mathbf{E}^{-1}(\hat{\boldsymbol{\omega}}^2 - \mathbf{T}\boldsymbol{\omega}^2)$$

$$- \sum_{i=1}^{m}(\log\tau_i - \tau_i\rho_i) - \frac{sqm}{2}\log\eta + \frac{\eta}{2}\left\|\hat{\boldsymbol{\psi}} - \boldsymbol{\Gamma}\boldsymbol{\phi}\right\|^2 - \log\nu + \nu\eta$$

$$+ \frac{1}{2}(\hat{\boldsymbol{\theta}}_u - \boldsymbol{\theta})^T \tilde{\mathbf{A}}^{-1}(\hat{\boldsymbol{\theta}}_u - \boldsymbol{\theta}) - \frac{dm}{2}\log\beta + \frac{\beta}{2}\sum_{i}^{m}\left\|(\mathbf{K}(\boldsymbol{\theta}) - \omega_i^2\,\mathbf{M})\boldsymbol{\phi}_i\right\|^2 \tag{21}$$

where we have dropped constants that do not depend on $\boldsymbol{\xi}$ and $\boldsymbol{\theta}$. The logarithm function in (21) is not a quadratic function for the whole uncertain model parameter vector $[\boldsymbol{\xi}^T, \boldsymbol{\theta}^T]^T$ but it is quadratic for each of the uncertain parameters $\boldsymbol{\phi}$, $\boldsymbol{\omega}^2$, and $\boldsymbol{\theta}$ if the other two parameters are fixed. Therefore, explicit expressions can be obtained for iterative minimization of the objective function in (21) to update all of the parameters in $[\boldsymbol{\xi}^T, \boldsymbol{\theta}^T]^T$ successively. This strategy is similar to [18] but they fix $\beta, \rho$ and $\eta$ a priori and so $\tau$ and $\nu$ are not needed. However, good values of $\beta, \rho$ and $\eta$ are difficult to choose a priori.

By minimizing the function in (21) with respect to $\boldsymbol{\phi}$, $\eta$ and $\nu$ successively with all other parameters fixed at their MAP values, the MAP estimates $\tilde{\boldsymbol{\phi}}$, $\tilde{\eta}$ and $\tilde{\nu}$ are expressed as:

$$\tilde{\boldsymbol{\phi}} = \left(\tilde{\beta}\tilde{\mathbf{F}} + \tilde{\eta}\boldsymbol{\Gamma}^T\boldsymbol{\Gamma}\right)^{-1}\tilde{\eta}\boldsymbol{\Gamma}^T\hat{\boldsymbol{\psi}} \tag{22}$$

$$\tilde{\eta} = \frac{sqm}{2\tilde{\nu} + \|\hat{\boldsymbol{\psi}} - \boldsymbol{\Gamma}\tilde{\boldsymbol{\phi}}\|^2} \tag{23}$$

$$\tilde{\nu} = 1/\tilde{\eta} \tag{24}$$

where:

$$\tilde{\mathbf{F}} = \begin{bmatrix} \left(\mathbf{K}(\tilde{\boldsymbol{\theta}}) - \tilde{\omega}_1^2\mathbf{M}\right)^2 & \cdots & \mathbf{0} \\ \vdots & \ddots & \vdots \\ \mathbf{0} & \cdots & \left(\mathbf{K}(\tilde{\boldsymbol{\theta}}) - \tilde{\omega}_m^2\mathbf{M}\right)^2 \end{bmatrix}_{dm\times dm} \tag{25}$$

We substitute (24) into (23) and the optimal estimate of $\tilde{\eta}$ is:

$$\tilde{\eta} = \frac{sqm-2}{\|\hat{\boldsymbol{\psi}} - \boldsymbol{\Gamma}\tilde{\boldsymbol{\phi}}\|^2} \tag{26}$$

This is consistent with an iterative solution for $\tilde{\eta}$ by iterating between (23) and (24) until convergence.

Similarly, with all other parameters fixed at their MAP values, the MAP estimates $\tilde{\boldsymbol{\omega}}^2$, $\tilde{\boldsymbol{\rho}}$ and $\tilde{\boldsymbol{\tau}}$ can be found by setting the derivatives of (21) with respect to $\boldsymbol{\omega}^2$, $\boldsymbol{\rho}$ and $\boldsymbol{\tau}$ equal to zero, respectively:

$$\tilde{\boldsymbol{\omega}}^2 = \left(\tilde{\beta}\tilde{\mathbf{G}}^T\tilde{\mathbf{G}} + \mathbf{T}^T\tilde{\mathbf{E}}^{-1}\mathbf{T}\right)^{-1}\left(\tilde{\beta}\tilde{\mathbf{G}}^T\tilde{\mathbf{c}} + \mathbf{T}^T\tilde{\mathbf{E}}^{-1}\hat{\boldsymbol{\omega}}^2\right) \tag{27}$$

$$\tilde{\rho}_i = \frac{q}{2\tilde{\tau}_i + \sum_{r=1}^{q}\left(\hat{\omega}_{r,i}^2 - \tilde{\omega}_i^2\right)^2} \tag{28}$$

$$\tilde{\tau}_i = 1/\tilde{\rho}_i \tag{29}$$

where the matrix $\tilde{\mathbf{E}}$ is $\mathbf{E}$ in (9) evaluated at $\tilde{\rho}_i$ and:

$$\tilde{\mathbf{G}} = \begin{bmatrix} \mathbf{M}\tilde{\boldsymbol{\phi}}_1 & \cdots & 0 \\ \vdots & \ddots & \vdots \\ 0 & \cdots & \mathbf{M}\tilde{\boldsymbol{\phi}}_m \end{bmatrix}_{dm\times m} \tag{30}$$

$$\tilde{\mathbf{c}} = \begin{bmatrix} \left(\mathbf{K}_0 + \sum_{j=1}^{n}\tilde{\theta}_j\mathbf{K}_j\right)\tilde{\boldsymbol{\phi}}_1 \\ \vdots \\ \left(\mathbf{K}_0 + \sum_{j=1}^{n}\tilde{\theta}_j\mathbf{K}_j\right)\tilde{\boldsymbol{\phi}}_m \end{bmatrix}_{dm\times 1} = \begin{bmatrix} \mathbf{K}(\tilde{\boldsymbol{\theta}})\tilde{\boldsymbol{\phi}}_1 \\ \vdots \\ \mathbf{K}(\tilde{\boldsymbol{\theta}})\tilde{\boldsymbol{\phi}}_m \end{bmatrix}_{dm\times 1} \tag{31}$$

We solve for $\tilde{\rho}_i$ using (29) in (28):

$$\tilde{\rho}_i = \frac{q-2}{\sum_{r=1}^{q}\left(\hat{\omega}_{r,i}^2 - \tilde{\omega}_i^2\right)^2} \tag{32}$$

This is consistent with an iterative solution for $\tilde{\rho}_i$ by iterating between (28) and (29) until convergence.

Following the corresponding procedure for $\boldsymbol{\theta}$, we minimize (21) with respect to $\boldsymbol{\theta}$ with other parameters fixed to get the MAP estimate of structural model parameter $\tilde{\boldsymbol{\theta}}$:

$$\tilde{\boldsymbol{\theta}} = \left(\tilde{\beta}\tilde{\mathbf{H}}^T\tilde{\mathbf{H}} + \tilde{\mathbf{A}}^{-1}\right)^{-1}\left(\tilde{\beta}\tilde{\mathbf{H}}^T\tilde{\mathbf{b}} + \tilde{\mathbf{A}}^{-1}\hat{\boldsymbol{\theta}}_u\right) = \left(\tilde{\beta}\tilde{\mathbf{A}}\tilde{\mathbf{H}}^T\tilde{\mathbf{H}} + \mathbf{I}_n\right)^{-1}\left(\tilde{\beta}\tilde{\mathbf{A}}\tilde{\mathbf{H}}^T\tilde{\mathbf{b}} + \hat{\boldsymbol{\theta}}_u\right) \tag{33}$$

where the matrix $\tilde{\mathbf{H}}$ and vector $\tilde{\mathbf{b}}$ are those defined in (5) and (6), respectively, but evaluated at the MAP values, $\tilde{\boldsymbol{\phi}}$ and $\tilde{\boldsymbol{\omega}}^2$.

Finally, by minimizing (21) with respect to $\beta$ for given $a_0$ and $b_0$, we get:

$$\tilde{\beta} = \frac{dm + 2(a_0 - 1)}{2b_0 + \sum_i^m \left\|(\mathbf{K}(\tilde{\boldsymbol{\theta}}) - \tilde{\omega}_i^2\,\mathbf{M})\tilde{\boldsymbol{\phi}}_i\right\|^2} \tag{34}$$

The final MAP estimates $\tilde{\boldsymbol{\xi}}$ and $\tilde{\boldsymbol{\theta}}$ are given by performing a sequence of iterations, in which (22), (26) (27), (32), (33) and (34) are successively evaluated until some convergence criterion is satisfied. Notice that the MAP hyper-parameters $\tilde{\boldsymbol{\delta}}$ are involved explicitly only in (33) where $\tilde{\alpha}_1, \dots, \tilde{\alpha}_n$ appear on the diagonal of $\tilde{\mathbf{A}}$. The determination of $\tilde{\boldsymbol{\delta}}$ is described in Subsection 2.3.3.

*Posterior uncertainty quantification of* $\boldsymbol{\xi}$ *and* $\boldsymbol{\theta}$

The posterior uncertainty in $\boldsymbol{\theta}$ and $\boldsymbol{\xi} = [\beta, (\boldsymbol{\omega}^2)^T, \boldsymbol{\rho}^T, \boldsymbol{\tau}^T, \boldsymbol{\phi}^T, \eta, \nu]^T$ can be expressed as follows. Laplace's method approximates the posterior PDF $p\left(\boldsymbol{\xi}, \boldsymbol{\theta} | \tilde{\boldsymbol{\delta}}, \hat{\boldsymbol{\omega}}^2, \hat{\boldsymbol{\psi}}, \hat{\boldsymbol{\theta}}_u\right)$ by a Gaussian distribution with the mean at the MAP estimates $\left(\tilde{\boldsymbol{\xi}}, \tilde{\boldsymbol{\theta}}\right)$ and covariance matrix $\boldsymbol{\Sigma}\left(\boldsymbol{\xi}, \boldsymbol{\theta} | \tilde{\boldsymbol{\delta}}, \hat{\boldsymbol{\omega}}^2, \hat{\boldsymbol{\psi}}, \hat{\boldsymbol{\theta}}_u\right)$ which is equal to the inverse of the Hessian of the objective function $\mathrm{J}(\boldsymbol{\xi}, \boldsymbol{\theta})$ evaluated at the MAP values $\left(\tilde{\boldsymbol{\xi}}, \tilde{\boldsymbol{\theta}}\right)$ [14]. The covariance matrix is therefore estimated as:

$$\boldsymbol{\Sigma}\left(\boldsymbol{\xi}, \boldsymbol{\theta} | \tilde{\boldsymbol{\delta}}, \hat{\boldsymbol{\omega}}^2, \hat{\boldsymbol{\psi}}, \hat{\boldsymbol{\theta}}_u\right) \approx \begin{bmatrix} \boldsymbol{\mathcal{H}}^{(1,1)} & \boldsymbol{\mathcal{H}}^{(1,2)} \\ \boldsymbol{\mathcal{H}}^{(2,1)} & \boldsymbol{\mathcal{H}}^{(2,2)} \end{bmatrix}^{-1} \tag{35}$$

where the block precision matrices are given by:

$$\boldsymbol{\mathcal{H}}^{(1,1)} = \begin{bmatrix} (dm/2 - 1 + \tilde{a}_0)\tilde{\beta}^{-2} & \left(\tilde{\mathbf{G}}^T\tilde{\mathbf{G}}\tilde{\boldsymbol{\omega}}^2 - \tilde{\mathbf{G}}^T\tilde{\mathbf{c}}\right)^T & 0 & 0 \\ & \tilde{\beta}\tilde{\mathbf{G}}^T\tilde{\mathbf{G}} + \mathbf{T}^T\tilde{\mathbf{E}}^{-1}\mathbf{T} & \mathrm{diag}\left(q\tilde{\omega}_1^2 - \sum_{r=1}^q \hat{\omega}_{r,1}^2, \dots, q\tilde{\omega}_m^2 - \sum_{r=1}^q \hat{\omega}_{r,m}^2\right) & \mathbf{0} \\ & & \mathrm{diag}(q\tilde{\rho}_1^{-2}, \dots, q\tilde{\rho}_m^{-2})/2 & \mathbf{I}_m \\ sym & & & \mathrm{diag}(\tilde{\tau}_1^{-2}, \dots, \tilde{\tau}_m^{-2}) \end{bmatrix}_{(3m+1)\times(3m+1)} \tag{36}$$

$$\mathcal{H}^{(2,2)} = \begin{bmatrix} \tilde{\beta}\tilde{\mathbf{F}} + \tilde{\eta}\boldsymbol{\Gamma}^T\boldsymbol{\Gamma} & \boldsymbol{\Gamma}^T(\boldsymbol{\Gamma}\tilde{\boldsymbol{\phi}} - \hat{\boldsymbol{\psi}}) & 0 & \tilde{\beta}\mathbf{L}_3 \\ & sqm\,\tilde{\eta}^{-2}/2 & 1 & 0 \\ & & \tilde{\nu}^{-2} & 0 \\ sym & & & \tilde{\beta}\tilde{\mathbf{H}}^T\tilde{\mathbf{H}} + \tilde{\mathbf{A}}^{-1} \end{bmatrix}_{(dm+n+2)\times(dm+n+2)} \tag{37}$$

$$\mathcal{H}^{(1,2)} = \begin{bmatrix} \tilde{\boldsymbol{\phi}}^T\tilde{\mathbf{F}}^T & 0 & 0 & (\tilde{\mathbf{H}}^T\tilde{\mathbf{H}}\tilde{\boldsymbol{\theta}} - \tilde{\mathbf{H}}^T\tilde{\mathbf{b}})^T \\ -2\tilde{\beta}\mathbf{L}_1 & 0 & 0 & -\tilde{\beta}\mathbf{L}_2 \\ 0 \quad 0 & & & 0 \quad 0 \\ 0 \quad 0 & & & 0 \quad 0 \end{bmatrix}_{(m+3)\times(dm+n+2)}, \ \mathcal{H}^{(2,1)} = [\mathcal{H}^{(1,2)}]^T \tag{38}$$

$$\mathbf{L}_1 = \begin{bmatrix} \tilde{\boldsymbol{\phi}}_1^T\mathbf{M}(\tilde{\mathbf{K}} - \tilde{\omega}_1^2\,\mathbf{M}) & & \mathbf{0} \\ & \ddots & \\ \mathbf{0} & & \tilde{\boldsymbol{\phi}}_m^T\mathbf{M}(\tilde{\mathbf{K}} - \tilde{\omega}_m^2\,\mathbf{M}) \end{bmatrix}_{m\times dm} \tag{39}$$

$$\mathbf{L}_2 = \begin{bmatrix} \tilde{\boldsymbol{\phi}}_1^T\mathbf{M}\mathbf{K}_1\tilde{\boldsymbol{\phi}}_1 & \cdots & \tilde{\boldsymbol{\phi}}_1^T\mathbf{M}\mathbf{K}_n\tilde{\boldsymbol{\phi}}_1 \\ \vdots & \ddots & \vdots \\ \tilde{\boldsymbol{\phi}}_m^T\mathbf{M}\mathbf{K}_1\tilde{\boldsymbol{\phi}}_m & \cdots & \tilde{\boldsymbol{\phi}}_m^T\mathbf{M}\mathbf{K}_n\tilde{\boldsymbol{\phi}}_m \end{bmatrix}_{m\times n} \tag{40}$$

$$\mathbf{L}_3 = \begin{bmatrix} [(\tilde{\mathbf{K}} - \tilde{\omega}_1^2\,\mathbf{M})\mathbf{K}_1 + \mathbf{K}_1(\tilde{\mathbf{K}} - \tilde{\omega}_1^2\,\mathbf{M})]\tilde{\boldsymbol{\phi}}_1 & \cdots & [(\tilde{\mathbf{K}} - \tilde{\omega}_1^2\,\mathbf{M})\mathbf{K}_n + \mathbf{K}_n(\tilde{\mathbf{K}} - \tilde{\omega}_1^2\,\mathbf{M})]\tilde{\boldsymbol{\phi}}_1 \\ \vdots & \ddots & \vdots \\ [(\tilde{\mathbf{K}} - \tilde{\omega}_m^2\,\mathbf{M})\mathbf{K}_1 + \mathbf{K}_1(\tilde{\mathbf{K}} - \tilde{\omega}_1^2\,\mathbf{M})]\tilde{\boldsymbol{\phi}}_m & \cdots & [(\tilde{\mathbf{K}} - \tilde{\omega}_m^2\,\mathbf{M})\mathbf{K}_n + \mathbf{K}_n(\tilde{\mathbf{K}} - \tilde{\omega}_m^2\,\mathbf{M})]\tilde{\boldsymbol{\phi}}_m \end{bmatrix}_{dm\times n} \tag{41}$$

where $\tilde{\mathbf{K}}$= $\mathbf{K}(\tilde{\boldsymbol{\theta}})$. Given the joint Gaussian posterior PDF with mean $[\tilde{\boldsymbol{\xi}}^T, \tilde{\boldsymbol{\theta}}^T]^T$ and covariance matrix given in (35), the mean and covariance matrix for the marginal Gaussian distribution of each parameter can be readily expressed by partitioning the joint mean and covariance matrix respectively.

<u>*Remark 2.3:*</u> The above MAP estimates and uncertainty quantification strategy are similar to that in [18]. The differences are that the hyper-parameters $\eta, \nu, \boldsymbol{\rho}, \boldsymbol{\tau}$ and $\beta$ are also optimized and more than one data segment (q $\geq$ 1) is allowed.

*2.3.3 MAP estimates and uncertainty quantification for PDF $p(\boldsymbol{\delta}|\hat{\boldsymbol{\omega}}^2, \hat{\boldsymbol{\psi}}, \hat{\boldsymbol{\theta}}_u)$*

In the next step, we find the MAP estimate $\tilde{\boldsymbol{\delta}}$ of the hyper-parameter $\boldsymbol{\delta}$. As stated in (20), these MAP values in Laplace's method are employed to approximate $p(\boldsymbol{\xi}, \boldsymbol{\theta}|\hat{\boldsymbol{\omega}}^2, \hat{\boldsymbol{\psi}}, \hat{\boldsymbol{\theta}}_u)$ under the assumption that $p(\boldsymbol{\delta}|\hat{\boldsymbol{\omega}}^2, \hat{\boldsymbol{\psi}}, \hat{\boldsymbol{\theta}}_u)$ is sharply peaked around its mode at the MAP values.

To find the MAP value of hyper-parameters $\boldsymbol{\delta} = [\boldsymbol{\alpha}, \lambda, \zeta]$, we maximize:

$$p(\boldsymbol{\delta}|\hat{\boldsymbol{\omega}}^2, \hat{\boldsymbol{\psi}}, \hat{\boldsymbol{\theta}}_u) = \int p(\boldsymbol{\delta}, \boldsymbol{\xi}|\hat{\boldsymbol{\omega}}^2, \hat{\boldsymbol{\psi}}, \hat{\boldsymbol{\theta}}_u) d\boldsymbol{\xi} = \int p(\boldsymbol{\delta}|\boldsymbol{\xi}, \hat{\boldsymbol{\omega}}^2, \hat{\boldsymbol{\psi}}, \hat{\boldsymbol{\theta}}_u) p(\boldsymbol{\xi}|\hat{\boldsymbol{\omega}}^2, \hat{\boldsymbol{\psi}}, \hat{\boldsymbol{\theta}}_u) d\boldsymbol{\xi}$$

$$\approx p(\boldsymbol{\delta}|\hat{\boldsymbol{\xi}}, \hat{\boldsymbol{\omega}}^2, \hat{\boldsymbol{\psi}}, \hat{\boldsymbol{\theta}}_u) \propto p(\hat{\boldsymbol{\omega}}^2, \hat{\boldsymbol{\psi}}, \hat{\boldsymbol{\theta}}_u|\hat{\boldsymbol{\xi}}, \boldsymbol{\delta}) p(\boldsymbol{\delta}) \tag{42}$$

where we assume the posterior $p(\boldsymbol{\xi}|\hat{\boldsymbol{\omega}}^2, \hat{\boldsymbol{\psi}}, \hat{\boldsymbol{\theta}}_u)$ has a unique maximum at $\hat{\boldsymbol{\xi}} = \arg\max p(\boldsymbol{\xi}|\hat{\boldsymbol{\omega}}^2, \hat{\boldsymbol{\psi}}, \hat{\boldsymbol{\theta}}_u)$ and

we apply Laplace's asymptotic approximation to the integral. The last part of (42) comes from Bayes' Theorem by dropping the denominator, which is independent of $\boldsymbol{\delta}$, and by noting that $\boldsymbol{\delta}$ and $\boldsymbol{\xi}$ are independent a priori, so $p(\boldsymbol{\delta}|\hat{\boldsymbol{\xi}}) = p(\boldsymbol{\delta})$.

We approximate $\hat{\boldsymbol{\xi}}$ by the MAP estimates $\tilde{\boldsymbol{\xi}}$ obtained in Subsection 2.3.2. Taking $\hat{\boldsymbol{\xi}} = \tilde{\boldsymbol{\xi}}$, in (42), the evidence function $p(\hat{\boldsymbol{\omega}}^2, \hat{\boldsymbol{\psi}}, \hat{\boldsymbol{\theta}}_u|\hat{\boldsymbol{\xi}}, \boldsymbol{\delta})$ of the model class $\mathcal{M}(\boldsymbol{\delta})$ is given by:

$$p(\hat{\boldsymbol{\omega}}^2, \hat{\boldsymbol{\psi}}, \hat{\boldsymbol{\theta}}_u|\tilde{\boldsymbol{\xi}}, \boldsymbol{\delta}) = \int p(\hat{\boldsymbol{\omega}}^2, \hat{\boldsymbol{\psi}}, \hat{\boldsymbol{\theta}}_u|\boldsymbol{\theta}, \tilde{\boldsymbol{\xi}}, \boldsymbol{\delta})\, p(\boldsymbol{\theta}|\tilde{\boldsymbol{\xi}}, \boldsymbol{\delta}) d\boldsymbol{\theta}$$

$$= \int p(\hat{\boldsymbol{\omega}}^2|\tilde{\boldsymbol{\omega}}^2, \tilde{\rho}) p(\hat{\boldsymbol{\psi}}|\tilde{\boldsymbol{\phi}}, \tilde{\eta})\, p(\hat{\boldsymbol{\theta}}_u|\boldsymbol{\theta}, \boldsymbol{\alpha}) p(\boldsymbol{\theta}|\tilde{\boldsymbol{\omega}}^2, \tilde{\boldsymbol{\phi}}, \tilde{\beta}) d\boldsymbol{\theta}$$

$$\propto \int p(\hat{\boldsymbol{\theta}}_u|\boldsymbol{\theta}, \boldsymbol{\alpha}) p(\boldsymbol{\theta}|\tilde{\boldsymbol{\omega}}^2, \tilde{\boldsymbol{\phi}}, \tilde{\beta})\, d\boldsymbol{\theta} = p(\hat{\boldsymbol{\theta}}_u|\tilde{\boldsymbol{\omega}}^2, \tilde{\boldsymbol{\phi}}, \tilde{\beta}, \boldsymbol{\alpha}) \tag{43}$$

Substituting the Gaussian PDFs in (7) and (12) into the last equation and integrating analytically over $\boldsymbol{\theta}$:

$$p(\hat{\boldsymbol{\theta}}_u|\tilde{\boldsymbol{\omega}}^2, \tilde{\boldsymbol{\phi}}, \tilde{\beta}, \boldsymbol{\alpha}) = \mathcal{N}\left(\hat{\boldsymbol{\theta}}_u \middle| (\tilde{\mathbf{H}}^T\tilde{\mathbf{H}})^{-1}\tilde{\mathbf{H}}^T\tilde{\mathbf{b}}, \mathbf{D}\right) \tag{44}$$

where $\mathbf{D} = \mathbf{A} + (\tilde{\beta}\tilde{\mathbf{H}}^T\tilde{\mathbf{H}})^{-1}$ and the matrix $\tilde{\mathbf{H}}$ is defined by (5), but evaluated at the MAP estimates $\tilde{\boldsymbol{\phi}}$. We call $p(\hat{\boldsymbol{\theta}}_u|\tilde{\boldsymbol{\omega}}^2, \tilde{\boldsymbol{\phi}}, \tilde{\beta}, \boldsymbol{\alpha})$ in (44) the *pseudo-evidence* for the structural model provided by the pseudo data $\hat{\boldsymbol{\theta}}_u$. From (42), (43) and (44), to find the MAP values $\tilde{\boldsymbol{\delta}}$, we must maximize:

$$p(\hat{\boldsymbol{\theta}}_u|\tilde{\boldsymbol{\omega}}^2, \tilde{\boldsymbol{\phi}}, \tilde{\beta}, \boldsymbol{\alpha}) p(\boldsymbol{\alpha}|\lambda) p(\lambda|\zeta)$$

$$= \mathcal{N}\left(\hat{\boldsymbol{\theta}}_u \middle| (\tilde{\mathbf{H}}^T\tilde{\mathbf{H}})^{-1}\tilde{\mathbf{H}}^T\tilde{\mathbf{b}}, \mathbf{D}\right) \cdot \lambda^n \exp\left(-\lambda \textstyle\sum_{j=1}^n \alpha_j\right) \cdot \zeta \exp(-\zeta\lambda) \tag{45}$$

As shown in Appendix A, direct differentiation of the logarithm of (43) with respect to $\boldsymbol{\alpha}$ and setting the derivative equal to zero, leads to:

$$\tilde{\alpha}_j = \frac{-1 + \sqrt{1 + 8\lambda\left((\boldsymbol{\Sigma}_{\boldsymbol{\theta}})_{jj} + (\hat{\boldsymbol{\theta}}_u - \tilde{\boldsymbol{\theta}})_j^2\right)}}{4\lambda} \tag{46}$$

where $\boldsymbol{\Sigma}_{\boldsymbol{\theta}} = (\tilde{\beta}\mathbf{A}\tilde{\mathbf{H}}^T\tilde{\mathbf{H}} + \mathbf{I}_n)^{-1}\mathbf{A} = \mathbf{A}(\tilde{\beta}\tilde{\mathbf{H}}^T\tilde{\mathbf{H}}\mathbf{A} + \mathbf{I}_n)^{-1}$ is the covariance matrix for $\boldsymbol{\theta}$ conditional on $\hat{\boldsymbol{\xi}}$, which corresponds to the inverse of the block precision matrix in the bottom right corner of $\boldsymbol{\mathcal{H}}^{(2,2)}$ in (37). A key point to note is that many of the $\tilde{\alpha}_j$ approach zero during the optimization, which implies from (11) that their corresponding $\Delta\theta_j = \theta_j - \hat{\theta}_{j,u}$ have negligibly small values. This is a similar procedure to sparse Bayesian learning where redundant or irrelevant features are pruned away leading to a sparse explanatory subset [19, 25]. Here, the procedure suppresses the occurrence of false damage detections by reducing the posterior uncertainty of the stiffness scaling parameter $\boldsymbol{\theta}$. To avoid the occurrence of missed alarms, an appropriate MAP value of hyper-parameters $\lambda$ should be utilized.

The MAP estimates of the hyper-parameters $\lambda$ and $\zeta$ are also derived in Appendix A:

$$\tilde{\lambda} = n/\left(\sum_{j=1}^{n} \tilde{\alpha}_j + \tilde{\zeta}\right) \tag{47}$$

$$\tilde{\zeta} = 1/\tilde{\lambda} \tag{48}$$

We do not eliminate $\tilde{\zeta}$ to give one equation for $\tilde{\lambda}$ because it may happen that all $\tilde{\alpha}_j$ are temporarily zero during an iteration.

The posterior PDF $p\left(\boldsymbol{\delta}|\hat{\boldsymbol{\omega}}^2, \hat{\boldsymbol{\psi}}, \hat{\boldsymbol{\theta}}_u\right)$ can be well approximated by a Gaussian distribution $\boldsymbol{\mathcal{N}}\left(\boldsymbol{\delta}|\tilde{\boldsymbol{\delta}}, \boldsymbol{\mathcal{H}}(\tilde{\boldsymbol{\delta}})^{-1}\right)$ with mean $\tilde{\boldsymbol{\delta}}$ and covariance matrix $\boldsymbol{\mathcal{H}}(\tilde{\boldsymbol{\delta}})^{-1}$[14], where the Hessian matrix $\boldsymbol{\mathcal{H}}(\tilde{\boldsymbol{\delta}})$ is calculated as:

$$\boldsymbol{\mathcal{H}}(\tilde{\boldsymbol{\delta}}) = \begin{bmatrix} 2\tilde{\mathbf{A}}^{-3}\tilde{\mathbf{B}} - \tilde{\mathbf{A}}^{-2} & \begin{bmatrix}1\\ \vdots \\ 1\end{bmatrix} & \begin{bmatrix}0\\ \vdots \\ 0\end{bmatrix} \\ \begin{bmatrix}1\\ \vdots \\ 1\end{bmatrix}^T & n\tilde{\lambda}^{-2} & 1 \\ \begin{bmatrix}0\\ \vdots \\ 0\end{bmatrix}^T & 1 & \tilde{\zeta}^{-2} \end{bmatrix}_{(n+2)\times(n+2)} \tag{49}$$

$$\text{where } \tilde{\mathbf{B}} = \begin{bmatrix} (\boldsymbol{\Sigma}_{\boldsymbol{\theta}})_{11} + \left(\hat{\boldsymbol{\theta}}_u - \tilde{\boldsymbol{\theta}}\right)_1^2 & \cdots & 0 \\ \vdots & \ddots & \vdots \\ 0 & \cdots & (\boldsymbol{\Sigma}_{\boldsymbol{\theta}})_{nn} + \left(\hat{\boldsymbol{\theta}}_u - \tilde{\boldsymbol{\theta}}\right)_n^2 \end{bmatrix}_{n\times n} \tag{50}$$

## 3. Proposed damage inference method

### *3.1 Sparse Bayesian learning algorithm for identification of spatially sparse stiffness reductions*

We produce a new Bayesian learning algorithm for sparse stiffness loss inference which iterates among the MAP values of all uncertain parameters until some specified convergence criteria are satisfied. Given the modal data $\hat{\boldsymbol{\omega}}^2$ and $\hat{\boldsymbol{\psi}}$, as well as the MAP estimates $\hat{\boldsymbol{\theta}}_u$ obtained from the calibration stage, the MAP estimates of the stiffness scaling parameters $\boldsymbol{\theta}$ are obtained, along with their corresponding posterior uncertainty. There are two variants of the algorithm. For the monitoring stage, Algorithm 2 is used that employs the evidence strategy in (46-48) for sparse stiffness loss inference. For the calibration stage, however, model sparseness is not expected and hence Algorithm 1 is used without optimization of the hyper-parameters $\{\boldsymbol{\alpha}, \lambda, \zeta\}$, so we fix all components $\alpha_j$ and $\lambda, \zeta$ are not needed.

**Algorithm 1 and 2:** Sparse Bayesian learning for identification of sparse stiffness reductions

1. **INPUTS:** Identified modal data $\hat{\boldsymbol{\omega}}^2$ and $\hat{\boldsymbol{\psi}}$, and if Algorithm 2, the MAP estimate $\hat{\boldsymbol{\theta}}_u$ from the calibration stage, while if Algorithm 1, a chosen nominal vector $\boldsymbol{\theta}_0$.
2. **OUTPUTS:** MAP estimates and posterior uncertainty of all uncertain parameters.
3. Initialize the stiffness scaling parameter vector with the nominal vector $\tilde{\boldsymbol{\theta}} = \boldsymbol{\theta}_0$ if the calibration stage (Algorithm 1) and if the monitoring stage, $\tilde{\boldsymbol{\theta}} = \hat{\boldsymbol{\theta}}_u$, and initialize the system natural frequencies as the mean of the measured natural frequencies over all data segments, $\tilde{\boldsymbol{\omega}}^2 = \sum_{r=1}^{q} \hat{\boldsymbol{\omega}}_r^2 / q$.
4. **While** convergence criterion is not met
5. Update MAP $\tilde{\boldsymbol{\phi}}$ using (22), and then update $\tilde{\eta}$ using (26);
6. Update MAP $\tilde{\boldsymbol{\omega}}^2$ using (27), and then update $\tilde{\boldsymbol{\rho}}$ using (32);
7. Update MAP $\tilde{\boldsymbol{\theta}}$ using (33);
8. Update MAP $\tilde{\beta}$ using (34) for given $a_0$ and $b_0$;
9. If Algorithm 1, fix all components in $\tilde{\boldsymbol{\alpha}}$ with large values (e.g. $\tilde{\alpha}_j = 10^9$);
10. If Algorithm 2, update $\tilde{\boldsymbol{\alpha}}$ using (46), then update $\tilde{\lambda}$ and $\tilde{\zeta}$ using (47) and (48);
11. **End while** (convergence criterion has been satisfied)
12. Estimate the posterior uncertainty using (35-41) and (49-50) for all uncertain parameters.

*3.3.1 Implementation details*

1. *Hyper-parameters initialization*. Using (26), (32) and (34) along with some approximations, the hyper-parameters $\beta$, $\eta$ and $\rho_i$ are initialized as:

$$\bar{\beta} = \big(dm + 2(a_0 - 1)\big)/2b_0, \tag{51}$$

$$\bar{\eta} = (sqm - 2)/\|\hat{\boldsymbol{\psi}}\|^2, \tag{52}$$

$$\bar{\rho}_i = \ (q-2)/\textstyle\sum_{r=1}^{q} \hat{\omega}_{r,i}^4\, , i = 1, \dots, m. \tag{53}$$

For the initial value of the hyper-parameters $\bar{\alpha}_j$, we use $\bar{\alpha}_j = n^2, j = 1, \dots, n$, which is inspired by [19].

2. *Optimization of the equation-error precision $\beta$.* An important step in the algorithm is the optimization of the equation-error precision $\beta$, where we fix the shape parameter $a_0$=1, which is chosen to produce what is considered a reasonable shape for penalizing values of $\beta$ that are too large, and the rate parameter $b_0$=1 and $b_0$=0.1, which is found to scale the prior distribution in (8) appropriately for Algorithms 1 and 2, respectively.

3. *Parameter fixing.* When running Algorithm 2, we fix components $\tilde{\theta}_j = \hat{\theta}_{u,j}$ if the corresponding hyper-parameter $\alpha_j$ becomes smaller than $\alpha_{min}$ (chosen as $10^{-9}$ in the example later), because this helps to accelerate the convergence of the algorithmic optimization.

4. *Convergence criterion.* Algorithm 2 is terminated when the change in all $\log \alpha_j's$ between the $l^{th}$ iteration and the $(l-1)^{th}$ iteration are sufficiently small (e.g. smaller than $0.005$). For Algorithm 1, the convergence criterion is satisfied when the change of model parameters in $\tilde{\boldsymbol{\theta}}$ is sufficiently small (e.g. smaller than 0.001).

5. *Number of data segments utilized.* It is seen from (32) that the number of data-segments for modal parameter identification should be at least three ($q \geq 3$) for tractable estimations of hyper-parameters $\tilde{\boldsymbol{\rho}}$.

<u>*Remark 3.1:*</u> Algorithm 2 is performed by iterating between the optimization of two groups of parameters $[\boldsymbol{\xi}^T, \boldsymbol{\theta}^T]^T$ and $\boldsymbol{\delta}$, which can be regarded as a generalized version of the Top-down scheme from sparse Bayesian learning [19,25]. The procedure starts by considering all substructures that are possibly damaged ($\tilde{\alpha}_j = n^2$ at the first iteration, $j = 1, \ldots, n$) and then causes the "inactive" components $\theta_j$ to be exactly equal to $\hat{\theta}_{u,j}$ from the calibration stage when optimizing over the hyper-parameters $\alpha_j$, so that finally only a few "active" $\theta_j's$ are changed and their corresponding substructures are considered to be damaged.

### *3.2 Evaluation of damage probability*

In contrast to the deterministic sparse inversion algorithms, such as, linear programming in Basis Pursuit [32-33] and greedy algorithms in Orthogonal Matching Pursuit [34], that provide only a point estimate of the model parameter vector to specify the sparse representation, the sparse Bayesian learning framework uses posterior probability distributions over the model parameters as an efficient way to quantify uncertainty of the sparse model. For the structural damage identification problem of interest here, the MAP estimates and the posterior uncertainty of the stiffness scaling parameters can be used to quantify the probability that the $j^{th}$ substructure stiffness scaling parameter has been reduced by more than a specified fraction $f$ of the stiffness in the calibration stage. To proceed, we denote the stiffness scaling parameter of the $j^{th}$ substructure for the current possibly damaged state and initial undamaged state as $\theta_{d,j}$ and $\theta_{u,j}$, respectively. Using a Gaussian asymptotic approximation [37] for the integrals involved, as in [2],

$$P_j^{dam}(f) = P\left(\theta_{d,j} < (1-f)\theta_{u,j} | \widehat{\boldsymbol{\omega}}_u^2, \widehat{\boldsymbol{\Psi}}_u, \widehat{\boldsymbol{\omega}}_d^2, \widehat{\boldsymbol{\Psi}}_d\right)$$

$$= \int P\left(\theta_{d,j} < (1-f)\theta_{u,j} | \theta_{u,j}, \widehat{\boldsymbol{\omega}}_u^2, \widehat{\boldsymbol{\Psi}}_u, \widehat{\boldsymbol{\omega}}_d^2, \widehat{\boldsymbol{\Psi}}_d\right) p\left(\theta_{u,j} | \widehat{\boldsymbol{\omega}}_u^2, \widehat{\boldsymbol{\Psi}}_u\right) d\theta_{u,j}$$

$$\approx \Phi\left[\frac{(1-f)\tilde{\theta}_{u,j} - \tilde{\theta}_{d,j}}{\sqrt{(1-f)^2\sigma_{d,j}^2 + \sigma_{u,j}^2}}\right] \tag{54}$$

where $\Phi(\cdot)$ is the standard Gaussian cumulative distribution function; $\left[\widehat{\boldsymbol{\omega}}_d^2, \widehat{\boldsymbol{\Psi}}_d\right]$ and $\left[\widehat{\boldsymbol{\omega}}_u^2, \widehat{\boldsymbol{\Psi}}_u\right]$ are the identified modal parameters from the current monitoring (possibly damaged) stage and calibration (initial undamaged) stage, respectively; $\tilde{\theta}_{d,j}$ and $\tilde{\theta}_{u,j} (= \hat{\theta}_{u,j})$ denote the MAP stiffness scaling parameters of the $j^{th}$ substructure for the possibly damaged and undamaged structure, respectively, from (33); $\sigma_{d,j}$ and $\sigma_{u,j}$ are the corresponding posterior standard deviations of the stiffness scaling parameter $\theta_j$ of the $j^{th}$ substructure, which are the square root of the diagonal elements of the posterior covariance matrix $\boldsymbol{\Sigma}_{\boldsymbol{\theta}}$ given after (46).

## 4 Examples

### *4.1 Example 1: Simulated ten-story shear building*

The first example is chosen to be the same as that presented in [18] and applies only Algorithm 1 for calibration. The structure is a ten-story linear shear building in which the lumped masses of each floor are equal to 100 metric tons. The inter-story stiffness is $k_0 = 176.729$MN/m for all stories to give the first five modal natural frequencies as 1.00, 2.98, 4.89, 6.69 and 8.34Hz. For system identification, one stiffness scaling parameter $\theta_j$ is used for each story, $j = 1, \ldots, 10$, where $\mathbf{K}_j = \theta_j \overline{\mathbf{K}}_j$ is the uncertain contribution of the $j^{th}$ story to the global stiffness matrix $\mathbf{K}$, as in (1) with $\mathbf{K}_0 = \mathbf{0}$, and $\overline{\mathbf{K}}_j$ is the 'nominal' contribution, which, for convenience, is taken as the exact contribution so $\theta_j = 1$ gives the exact value for $\mathbf{K}_j$ at the calibration (undamaged) stage. Zero-mean Gaussian noise is added to the exact modal parameters with a standard deviation of 1% of the exact modal frequencies and mode shapes, which is the same strategy as in the method by Yuen et al. [18]. The goal of this example is to test the performance of Algorithm 1 for identifying the structural model parameters $\boldsymbol{\theta}$ at the calibration stage. As in [18], the initial value of each $\theta_j$ for starting the iterations in Algorithm 1 is selected randomly from a uniform distribution within the interval between 2 and 3 to demonstrate robustness to the initial choice.

In the first set of experiments, we study the effect of different choices of the equation-error precision $\beta$ on the identification performance. First, the results of the identified MAP values of the stiffness scaling parameters $\theta_j$ using four choices of hyper-parameter $\beta$ and the first four measured modes $(m = 4)$ identified from three data segments ($q = 3$) of complete measurements ($s = 10$) are tabulated in Table 1 for the method by Yuen et al.[18]. The associated posterior coefficients of variation (c.o.v), which are calculated from the ratio of the square root of the diagonal elements of the posterior covariance matrix $\boldsymbol{\Sigma}_{\boldsymbol{\theta}}$ given after (46) to the MAP values in (33), are also presented in Table 1. In order to make a fair comparison with the results in [18], we set extra parameters $\varphi_i = \rho_i \sum_{r=1}^{q} \hat{\omega}_{r,i}^4 / q, i = 1, \dots, m,$ to normalize $\rho_i$ and the corresponding parameter vector is $\boldsymbol{\varphi} = [\varphi_1, \dots, \varphi_m]$. Following the strategy used in [18], we fix the hyper-parameters $\eta = 10^5$ and $\boldsymbol{\varphi} = [10^4, \dots, 10^4]$. It is seen that the various values of $\beta$ correspond to different MAP identification results for the method by Yuen et al. [18] and hence proper selection of the hyper-parameter $\beta$ is important for identification accuracy. It is further observed that the associated posterior uncertainty is highly dependent on the selected value of $\beta$, with coefficients of variation that can give a misleading confidence in the accuracy of the MAP estimates.

Next, we run Algorithm 1 with different choices of the initial value of $\beta$ and with fixed hyper-parameters $\eta = 10^5$ and $\boldsymbol{\varphi} = [10^4, \dots, 10^4]$ using from three to five measured modes $(m)$ identified from three data segments ($q = 3$). For each set of measured modes, all runs converge to the same MAP value of $\beta$ (bottom line of Table 2) and the same MAP vector and associated c.o.v of the stiffness scaling parameters $\boldsymbol{\theta}$, no matter what initial values of β is chosen. The identification error for the obtained results for four measured modes is generally smaller than those for the method by Yuen et al. [18] that are shown in Table 1. Therefore, the use of the optimization scheme for the selection of the hyper-parameters $\beta$ in Algorithm 1 gives it the ability to accurately identify the stiffness scaling parameters $\boldsymbol{\theta}$. Comparing the associated posterior uncertainty, the c.o.v values obtained for Algorithm 1 lie between those obtained by the method by Yuen et al. [18] for $\beta = 0.1\bar{\beta}$ and $\bar{\beta}$ due to the final MAP estimate $\tilde{\beta}$=17.643, where $\bar{\beta} = 20$ is calculated using (51) with $a_0$=1 and $b_0$=1. The results in Table 2 shows, as expected, that using more measured modes results in smaller identification errors and smaller associated uncertainty than the results when using less modes; while the identified MAP value $\tilde{\beta}$ gets larger.

The iteration histories for convergence of the MAP values of the stiffness scaling parameters are shown in Figure 2 (a) and (b) corresponding to the results in Table 1 and 2. When the selected $\beta$ is small, i.e., $\beta = 0.1\bar{\beta}$, the convergence for the method by Yuen et al. [18] is very fast, occurring in the first few iterations, but the final identified MAP values have larger errors than in Algorithm 1. Larger $\beta$ should be selected for more accurate identification results. However, when $\beta$ is selected to be too large, i.e., $\beta = 100\bar{\beta}$, the convergence is very slow, requiring more than 300 iterations. In contrast, Algorithm 1 produces accurate identification results in the first few iterations, no matter what the initial $\beta$ is, as shown in Figure 2 (b), which shows an advantage for employing a hierarchical Bayesian prior for $\beta$ in (8) and then finding the MAP value of $\beta$ from the modal data.

Finally, we examine optimization of the hyper-parameters $\eta$ and $\boldsymbol{\rho}$ for good identification performance. We run Algorithm 1 to optimize the hyper-parameters $\beta, \eta$ and $\boldsymbol{\rho}$ and vary the $\bar{\beta}$, $\bar{\eta}$, and $\bar{\varphi}_i = \bar{\rho}_i \sum_{r=1}^{q} \hat{\omega}_{r,i}^4 / q$ calculated from (51), (52) and (53), respectively, by factors 0.1,1,10 and 100 to get four choices of the initial values, using from three to five measured modes ($m = 3 - 5$) identified from three data segments ($q = 3$). Identical results are obtained for all runs and the final identified MAP values and their associated c.o.v are presented in Table 3. The identification accuracy of the MAP values and the associated c.o.v are close to that in Table 2, because of the similar MAP estimates $\tilde{\beta}$. Furthermore, it is observed that the identified MAP estimates $\tilde{\eta}$ and $\tilde{\varphi}_i = \tilde{\rho}_i \sum_{r=1}^{q} \hat{\omega}_{r,i}^4 / q$ , $i = 1, \dots m$, are close to the values selected in the method by Yuen et al. [18], i.e., $10^5$ and $10^4$, respectively. Figure 2 (c) shows the iteration history for convergence of the MAP estimates $\tilde{\boldsymbol{\theta}}$ and the values essentially converge in 80 iterations, which is much slower convergence than in Figure 2 (b). This is due to the additional optimization of the hyper-parameters $\eta$ and $\boldsymbol{\rho}$.

**Table 1**

Identification results using the method by Yuen et al. [18] with three data segments ($q = 3$) and four measured modes ($m = 4$), and $\eta = 10^5$, $\boldsymbol{\varphi} = [10^4, \dots, 10^4]$ and $\beta$ fixed at different values (Example 1)

| $\beta$ | | $0.1\bar{\beta}$* | | $\bar{\beta}$ | | $10\bar{\beta}$ | | $100\bar{\beta}$ | |
|---|---|---|---|---|---|---|---|---|---|
| Para-meters | Initial values | MAP | c.o.v. (%) | MAP | c.o.v. (%) | MAP | c.o.v. (%) | MAP | c.o.v. (%) |
| $\theta_1$ | 2.033 | 0.990 | 0.871 | 0.990 | 0.275 | 0.990 | 0.087 | 0.989 | 0.028 |
| $\theta_2$ | 2.462 | 0.991 | 1.500 | 0.993 | 0.474 | 0.992 | 0.150 | 0.995 | 0.047 |
| $\theta_3$ | 2.771 | 1.000 | 0.798 | 1.001 | 0.252 | 1.001 | 0.080 | 1.002 | 0.025 |
| $\theta_4$ | 2.268 | 1.002 | 0.715 | 1.000 | 0.226 | 0.999 | 0.071 | 0.996 | 0.023 |
| $\theta_5$ | 2.583 | 1.003 | 0.956 | 1.001 | 0.303 | 1.002 | 0.096 | 1.001 | 0.030 |
| $\theta_6$ | 2.936 | 0.987 | 0.736 | 0.991 | 0.233 | 0.995 | 0.074 | 0.997 | 0.023 |
| $\theta_7$ | 2.410 | 0.998 | 0.763 | 1.000 | 0.241 | 1.000 | 0.076 | 1.001 | 0.024 |
| $\theta_8$ | 2.348 | 1.005 | 0.999 | 0.998 | 0.317 | 0.993 | 0.100 | 0.997 | 0.032 |
| $\theta_9$ | 2.148 | 0.992 | 0.683 | 0.993 | 0.216 | 0.992 | 0.068 | 0.987 | 0.022 |
| $\theta_{10}$ | 2.305 | 0.998 | 0.750 | 0.998 | 0.237 | 0.998 | 0.075 | 1.001 | 0.024 |

*$\bar{\beta} = dm/2 = 20$ calculated using (51) with $a_0$=1 and $b_0$=1.

**Table 2**

Identification results of Algorithm 1 using different number of measured modes ($m$) but the same number of date segments ($q = 3$), with $\eta = 10^5$, $\boldsymbol{\varphi} = [10^4, \dots, 10^4]$ and $\beta$ optimized from different initial values (Example 1)

| Para meter | Initial values | 3 modes ($m = 3$) | | 4 modes ($m = 4$) | | 5 modes ($m = 5$) | |
|---|---|---|---|---|---|---|---|
| | | MAP | c.o.v. (%) | MAP | c.o.v. (%) | MAP | c.o.v. (%) |
| $\theta_1$ | 2.033 | 0.995 | 0.573 | 0.990 | 0.293 | 0.995 | 0.178 |
| $\theta_2$ | 2.462 | 0.983 | 0.677 | 0.993 | 0.504 | 0.997 | 0.278 |
| $\theta_3$ | 2.771 | 1.018 | 0.844 | 1.001 | 0.268 | 0.996 | 0.139 |
| $\theta_4$ | 2.268 | 0.993 | 0.498 | 1.001 | 0.241 | 0.999 | 0.189 |
| $\theta_5$ | 2.583 | 1.017 | 0.455 | 1.001 | 0.322 | 1.000 | 0.144 |
| $\theta_6$ | 2.936 | 0.993 | 0.548 | 0.992 | 0.248 | 0.998 | 0.176 |
| $\theta_7$ | 2.410 | 0.994 | 0.638 | 0.999 | 0.256 | 0.995 | 0.151 |
| $\theta_8$ | 2.348 | 0.992 | 0.468 | 0.999 | 0.337 | 0.995 | 0.171 |
| $\theta_9$ | 2.148 | 1.000 | 0.440 | 0.993 | 0.230 | 0.993 | 0.165 |
| $\theta_{10}$ | 2.305 | 1.005 | 0.564 | 0.998 | 0.252 | 0.997 | 0.136 |
| $\beta$ | $\{0.1,1,10,100\} \times \bar{\beta}$* | 14.339 | 25.820 | 17.643 | 22.361 | 22.768 | 20.000 |

*$\bar{\beta} = dm/2 = 5m$, calculated using (51) with $a_0$=1 and $b_0$=1.

**Table 3**

Identification results of Algorithm 1 using different number of measured modes ($m$) but the same number of date segments ($q = 3$), with $\beta,\ \eta$ and $\boldsymbol{\rho}$ optimized from different initial values (Example 1)

| Parameter | Initial values | 3 modes ($m = 3$) MAP | 3 modes ($m = 3$) c.o.v. (%) | 4 modes ($m = 4$) MAP | 4 modes ($m = 4$) c.o.v. (%) | 5 modes ($m = 5$) MAP | 5 modes ($m = 5$) c.o.v. (%) |
|---|---|---|---|---|---|---|---|
| $\theta_1$ | 2.033 | 0.996 | 0.579 | 0.990 | 0.298 | 0.995 | 0.179 |
| $\theta_2$ | 2.462 | 0.984 | 0.684 | 0.993 | 0.512 | 0.997 | 0.281 |
| $\theta_3$ | 2.771 | 1.018 | 0.854 | 1.002 | 0.273 | 0.997 | 0.140 |
| $\theta_4$ | 2.268 | 0.993 | 0.504 | 1.001 | 0.244 | 0.999 | 0.190 |
| $\theta_5$ | 2.583 | 1.017 | 0.461 | 1.001 | 0.327 | 1.000 | 0.146 |
| $\theta_6$ | 2.936 | 0.994 | 0.554 | 0.991 | 0.251 | 0.998 | 0.178 |
| $\theta_7$ | 2.410 | 0.995 | 0.646 | 1.000 | 0.260 | 0.995 | 0.152 |
| $\theta_8$ | 2.348 | 0.992 | 0.473 | 0.999 | 0.342 | 0.995 | 0.172 |
| $\theta_9$ | 2.148 | 1.001 | 0.445 | 0.993 | 0.234 | 0.994 | 0.167 |
| $\theta_{10}$ | 2.305 | 1.006 | 0.571 | 0.998 | 0.256 | 0.997 | 0.137 |
| $\beta$ | $\{0.1,1,10,100\} \times \bar{\beta}$* | 14.000 | 25.820 | 17.099 | 22.361 | 22.349 | 20.000 |
| $\eta$ | $\{0.1,1,10,100\} \times \bar{\eta}$* | $1.397\times 10^5$ | 14.907 | $1.255\times 10^5$ | 12.910 | $1.146\times 10^5$ | 11.547 |
| $\varphi_1$ | $\{0.1,1,10,100\} \times \bar{\varphi}_1$* | $0.318\times 10^4$ | 81.650 | $0.318\times 10^4$ | 81.650 | $0.318\times 10^4$ | 81.650 |
| $\varphi_2$ | $\{0.1,1,10,100\} \times \bar{\varphi}_2$* | $0.350\times 10^4$ | 81.650 | $0.351\times 10^4$ | 81.650 | $0.340\times 10^4$ | 81.650 |
| $\varphi_3$ | $\{0.1,1,10,100\} \times \bar{\varphi}_3$* | $0.977\times 10^4$ | 81.650 | $0.578\times 10^4$ | 81.650 | $0.627\times 10^4$ | 81.650 |
| $\varphi_4$ | $\{0.1,1,10,100\} \times \bar{\varphi}_4$* | | | $0.552\times 10^4$ | 81.650 | $0.552\times 10^4$ | 81.650 |
| $\varphi_5$ | $\{0.1,1,10,100\} \times \bar{\varphi}_5$* | | | | | $0.215\times 10^4$ | 81.650 |

*$\bar{\beta} = dm/2 = 5m$, calculated using (51) with $a_0$=1 and $b_0$=1;

*$\bar{\eta} = (sqm - 2)/\|\hat{\boldsymbol{\psi}}\|^2 = 30m - 2$ (with normalization $\|\hat{\boldsymbol{\psi}}\| = 1$) calculated from (52);

* $\bar{\varphi}_i = \bar{\rho}_i \sum_{r=1}^{q} \hat{\omega}_{r,i}^4 / q = (q-2)/q = 1/3$ calculated from (53).

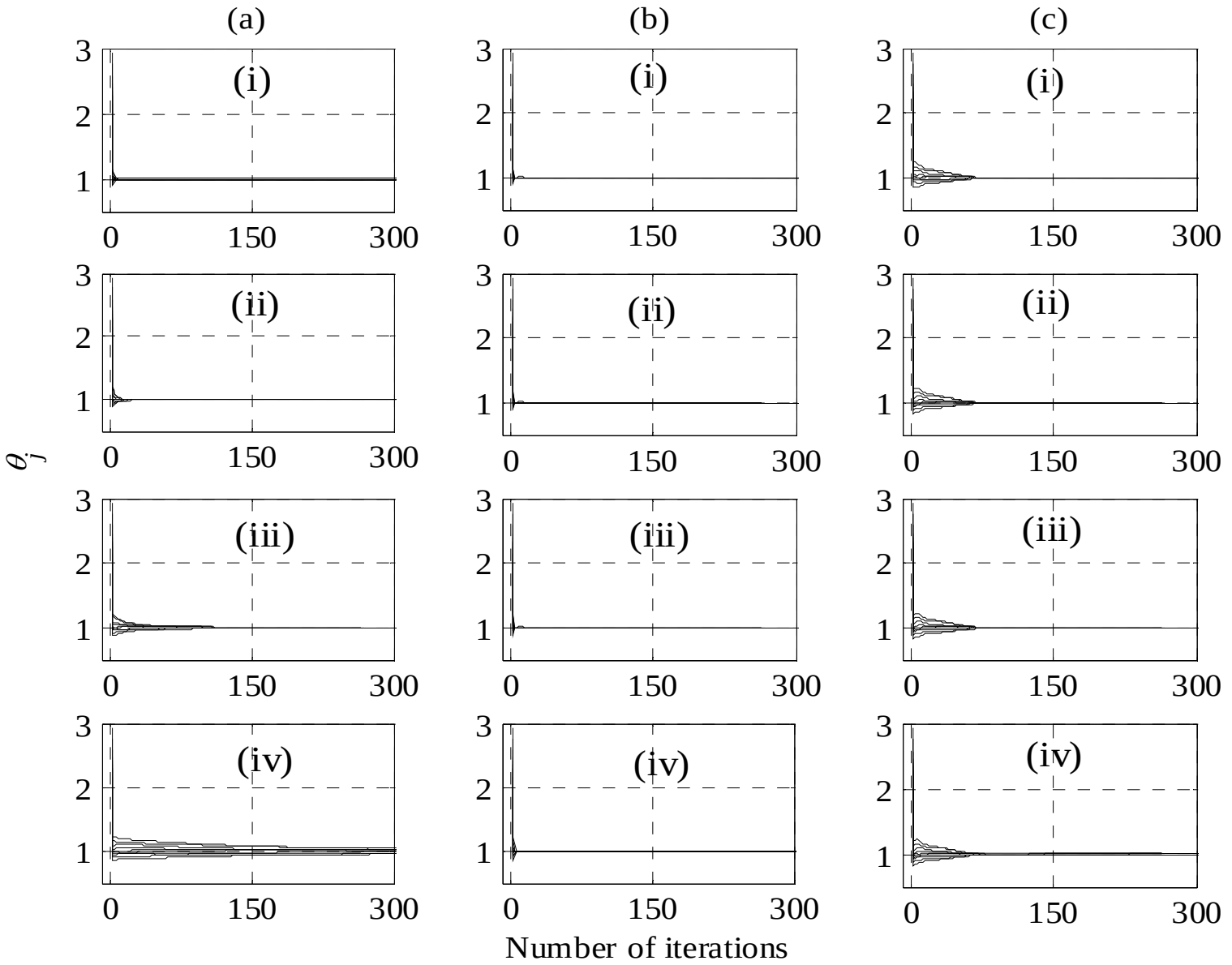

**Fig. 2.** Iteration histories for the MAP values of the 10 stiffness scaling parameter ($m = 4, q = 3$): (a) using the method by Yuen et al. [18] with $\eta = 10^5$, $\boldsymbol{\varphi} = [10^4, \ldots, 10^4]$ and $\beta$ fixed at different values: (i)$0.1\bar{\beta}$; (ii) $\bar{\beta}$; (iii) $10\bar{\beta}$; (iv) $100\bar{\beta}$; (b) using Algorithm 1 with $\eta = 10^5, \boldsymbol{\varphi} = [10^4, \ldots, 10^4]$ and $\beta$ optimized starting from different initial values: (i)$0.1\bar{\beta}$; (ii) $\bar{\beta}$; (iii) $10\bar{\beta}$; (iv) $100\bar{\beta}$; (c) using Algorithm 1 with $\beta, \eta$ and $\rho_i$ optimized and varying the initial default values $\bar{\beta}, \bar{\eta}$ and $\bar{\boldsymbol{\varphi}}$ by different factors: (i) 0.1; (ii) 1; (iii) 10; (iv) 100.

We now consider different numbers of data segments to give multiple estimates of the identified modal parameters. The stiffness identification results using Algorithm 1 are presented in Tables 4 and 5, for full-sensor and partial-sensor scenarios, respectively. For the full-sensor scenario, measurements at all ten floors are available ($s = 10$) while for the partial-sensor scenario, only measurements from five sensors are utilized which are located on the first, fourth, fifth, seventh and top floors ($s = 5$). The corresponding iteration histories for the MAP values are also shown in Figure 3. For stiffness parameter inference, the first four measured modes ($m = 4$) are utilized and the initial values of the hyper-parameters are set as $\bar{\beta}, \bar{\eta}$ and $\bar{\boldsymbol{\varphi}} = [\bar{\varphi}_1, \dots, \bar{\varphi}_i]$, calculated from (51), (52) and (53), respectively. It is not surprising to see that the MAP estimates $\tilde{\boldsymbol{\theta}}$ become closer and closer to their actual value with the increase of the number of the data segments. Moreover, the corresponding uncertainty also decreases, implying higher confidence in the identification results. Correspondingly, much fewer numbers of iterations are required for convergences, as seen from the observation of the iteration histories in Figure 3. All of these benefits come from more information in the data that is available for constraining the parameter updating. Notice that when the number of data segments is selected to be 100, the identification errors for the MAP estimates of all stiffness scaling parameters $\theta_j$ become smaller than 0.3%, which is accurate enough for the MAP values $\tilde{\boldsymbol{\theta}}$ to be utilized as the pseudo data for the likelihood function in (12) in a subsequent monitoring stage. The convergence of the MAP values occurs in about 30 iterations for the full-sensor scenario, which is a much smaller number than that of the tests when only three data segments is utilized (compare Figure 3 with Figure 2).

The final six rows of Tables 4 and 5 give the identified MAP estimates $\tilde{\beta}$, $\tilde{\eta}$ and $\tilde{\boldsymbol{\varphi}}$. It is seen the identified $\tilde{\beta}'s$ are around 18 and 19 for full-sensor and partial-sensor scenarios, respectively, which is close to the results in Table 3. The identified MAP estimates $\tilde{\eta}$ and $\tilde{\varphi}_i$ get closer and closer to the values selected in the method by Yuen et al. [18], i.e., $10^5$ and $10^4$, respectively, with the increase of the number of the data segments. However, the proposed algorithm is automatic and no user-intervention is needed, which is an advantage of employing a hierarchical Bayesian prior for $\eta$ and $\boldsymbol{\rho}$ in (10) and finding their MAP values from the data.

**Table 4**

Identification results by Algorithm 1 using various number of data segments for the full-sensor scenario ($s = 10$) (Example 1)

| Para-meter | Initial values | 5 segments MAP | 5 segments c.o.v. (%) | 10 segments MAP | 10 segments c.o.v. (%) | 50 segments MAP | 50 segments c.o.v. (%) | 100 segments MAP | 100 segments c.o.v. (%) |
|---|---|---|---|---|---|---|---|---|---|
| $\theta_1$ | 2.033 | 0.986 | 0.304 | 0.997 | 0.293 | 0.999 | 0.284 | 1.000 | 0.284 |
| $\theta_2$ | 2.462 | 0.996 | 0.524 | 1.006 | 0.502 | 1.010 | 0.486 | 1.004 | 0.486 |
| $\theta_3$ | 2.771 | 0.996 | 0.279 | 0.993 | 0.269 | 1.000 | 0.261 | 1.000 | 0.261 |
| $\theta_4$ | 2.268 | 1.001 | 0.250 | 0.997 | 0.241 | 0.999 | 0.234 | 0.999 | 0.233 |
| $\theta_5$ | 2.583 | 0.996 | 0.335 | 0.999 | 0.322 | 0.999 | 0.313 | 1.001 | 0.312 |
| $\theta_6$ | 2.936 | 0.988 | 0.257 | 0.995 | 0.248 | 1.000 | 0.240 | 1.000 | 0.240 |
| $\theta_7$ | 2.410 | 0.993 | 0.266 | 0.992 | 0.256 | 0.999 | 0.248 | 1.000 | 0.248 |
| $\theta_8$ | 2.348 | 0.995 | 0.350 | 1.000 | 0.335 | 0.999 | 0.327 | 0.998 | 0.326 |
| $\theta_9$ | 2.148 | 0.988 | 0.238 | 1.000 | 0.229 | 0.999 | 0.223 | 1.000 | 0.223 |
| $\theta_{10}$ | 2.305 | 0.994 | 0.261 | 0.995 | 0.251 | 1.001 | 0.245 | 1.000 | 0.244 |
| $\beta$ | $\bar{\beta}$* | 16.523 | 22.361 | 17.779 | 22.361 | 18.699 | 22.361 | 18.786 | 22.361 |
| $\eta$ | $\bar{\eta}$* | $1.169\times 10^5$ | 10.000 | $1.225\times 10^5$ | 7.071 | $1.046\times 10^5$ | 3.162 | $1.012\times 10^5$ | 2.236 |
| $\varphi_1$ | $\bar{\varphi}_1$* | $0.492\times 10^4$ | 63.246 | $0.994\times 10^4$ | 44.721 | $1.151\times 10^4$ | 20.000 | $1.018\times 10^4$ | 14.142 |
| $\varphi_2$ | $\bar{\varphi}_2$* | $0.523\times 10^4$ | 63.246 | $0.591\times 10^4$ | 44.721 | $0.921\times 10^4$ | 20.000 | $0.957\times 10^4$ | 14.142 |
| $\varphi_3$ | $\bar{\varphi}_3$* | $0.489\times 10^4$ | 63.246 | $0.631\times 10^4$ | 44.721 | $0.905\times 10^4$ | 20.000 | $0.864\times 10^4$ | 14.142 |
| $\varphi_4$ | $\bar{\varphi}_4$* | $1.432\times 10^4$ | 63.246 | $1.676\times 10^4$ | 44.721 | $0.854\times 10^4$ | 20.000 | $0.939\times 10^4$ | 14.142 |

*$\bar{\beta} = dm/2 = 20$, calculated using (51) with $a_0$=1 and $b_0$=1;

*$\bar{\eta} = (sqm-2)/\|\hat{\boldsymbol{\Psi}}\|^2 = 40q - 2$ calculated from (52);

*$\bar{\varphi}_i = \bar{\rho}_i \sum_{r=1}^{q} \hat{\omega}_{r,i}^4 / q = (q-2)/q$, $i = 1, \dots, 4$, calculated from (53).

**Table 5**

Identification results by Algorithm 1 using various number of data segments for the partial-sensor scenario ($s = 5$) (Example 1)

| Para-meter | Initial values | 5 segments MAP | 5 segments c.o.v. (%) | 10 segments MAP | 10 segments c.o.v. (%) | 50 segments MAP | 50 segments c.o.v. (%) | 100 segments MAP | 100 segments c.o.v. (%) |
|---|---|---|---|---|---|---|---|---|---|
| $\theta_1$ | 2.033 | 0.990 | 0.284 | 0.999 | 0.282 | 0.997 | 0.279 | 0.999 | 0.276 |
| $\theta_2$ | 2.462 | 1.014 | 0.487 | 1.017 | 0.482 | 1.013 | 0.477 | 1.003 | 0.473 |
| $\theta_3$ | 2.771 | 0.994 | 0.261 | 0.994 | 0.260 | 1.002 | 0.256 | 1.002 | 0.254 |
| $\theta_4$ | 2.268 | 0.997 | 0.234 | 0.992 | 0.232 | 1.000 | 0.229 | 0.999 | 0.227 |
| $\theta_5$ | 2.583 | 0.998 | 0.313 | 0.997 | 0.311 | 0.998 | 0.307 | 1.002 | 0.304 |
| $\theta_6$ | 2.936 | 1.003 | 0.241 | 0.999 | 0.239 | 1.000 | 0.236 | 1.000 | 0.233 |
| $\theta_7$ | 2.410 | 0.985 | 0.246 | 0.995 | 0.246 | 0.998 | 0.243 | 0.999 | 0.241 |
| $\theta_8$ | 2.348 | 1.006 | 0.328 | 1.008 | 0.324 | 1.003 | 0.320 | 1.002 | 0.317 |
| $\theta_9$ | 2.148 | 0.979 | 0.222 | 0.992 | 0.221 | 0.996 | 0.218 | 0.998 | 0.216 |
| $\theta_{10}$ | 2.305 | 0.991 | 0.244 | 0.996 | 0.242 | 1.004 | 0.240 | 1.001 | 0.237 |
| $\beta$ | $\bar{\beta}$* | 18.922 | 22.361 | 19.132 | 22.361 | 19.502 | 22.361 | 19.832 | 22.361 |
| $\eta$ | $\bar{\eta}$* | $1.332\times 10^5$ | 14.142 | $1.290\times 10^5$ | 10.000 | $1.013\times 10^5$ | 4.472 | $1.015\times 10^5$ | 3.162 |
| $\varphi_1$ | $\bar{\varphi}_1$* | $0.492\times 10^4$ | 63.246 | $0.995\times 10^4$ | 44.721 | $1.153\times 10^4$ | 20.000 | $1.018\times 10^4$ | 14.142 |
| $\varphi_2$ | $\bar{\varphi}_2$* | $0.540\times 10^4$ | 63.246 | $0.593\times 10^4$ | 44.721 | $0.920\times 10^4$ | 20.000 | $0.957\times 10^4$ | 14.142 |
| $\varphi_3$ | $\bar{\varphi}_3$* | $0.499\times 10^4$ | 63.246 | $0.625\times 10^4$ | 44.721 | $0.905\times 10^4$ | 20.000 | $0.864\times 10^4$ | 14.142 |
| $\varphi_4$ | $\bar{\varphi}_4$* | $1.481\times 10^4$ | 63.246 | $1.681\times 10^4$ | 44.721 | $0.856\times 10^4$ | 20.000 | $0.939\times 10^4$ | 14.142 |

*$\bar{\beta} = dm/2 = 20$, calculated using (51) with $a_0$=1 and $b_0$=1;

*$\bar{\eta} = (sqm-2)/\|\hat{\boldsymbol{\Psi}}\|^2 = 20q - 2$ calculated from (52);

*$\bar{\varphi}_i = \bar{\rho}_i \sum_{r=1}^{q} \hat{\omega}_{r,i}^4 / q = (q-2)/q$, $i = 1, \dots, 4$, calculated from (53).

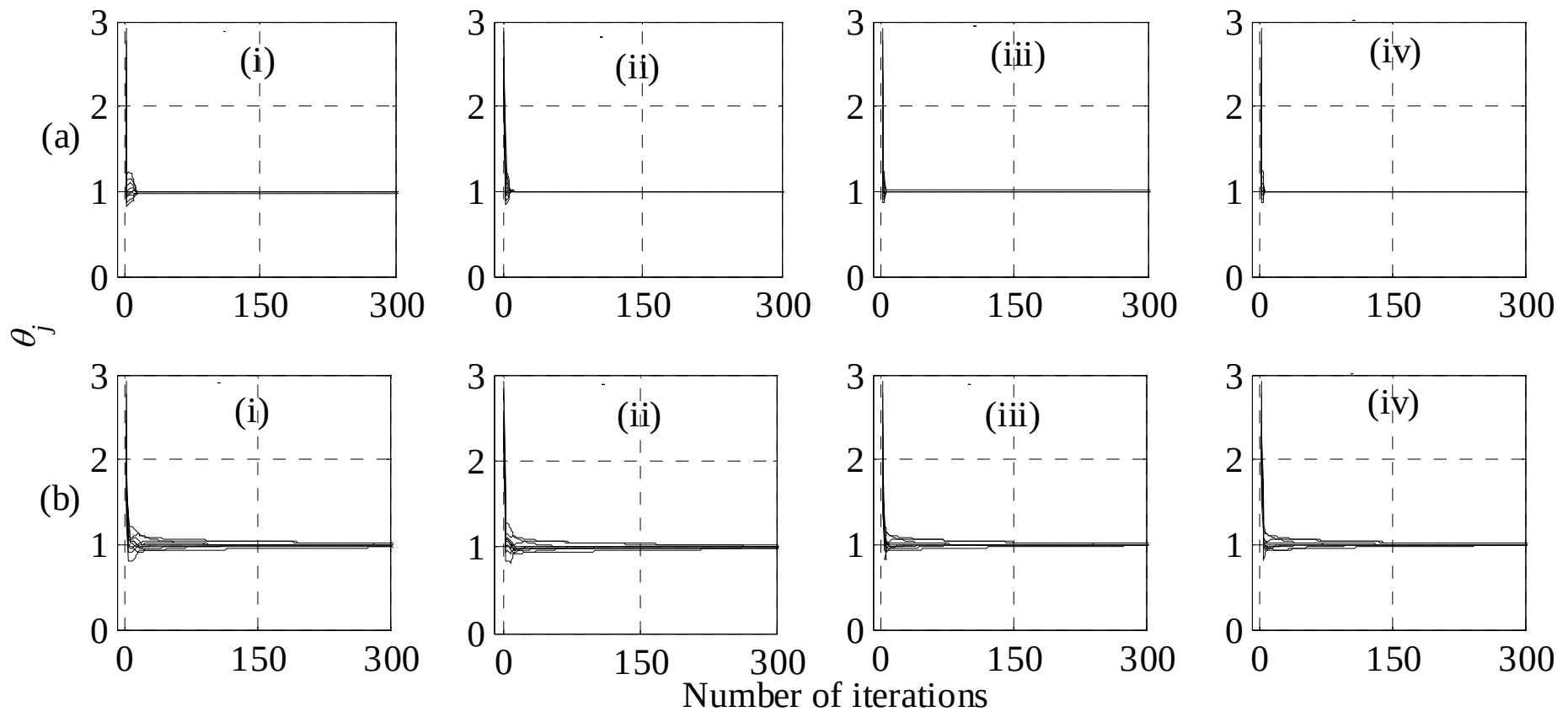

**Fig. 3.** Iteration histories for the MAP values of the 10 stiffness scaling parameter using Algorithm 1 for two sensor scenarios: (a) full-sensor; (b) partial-sensor; and with four different numbers of data segments utilized: (i) 5; (ii) 10; (iii) 50; (iv) 100.

### *4.2. Example 2: IASC-ASCE Phase II Simulation Benchmark problem*

#### *4.2.1. Structure and modal data*

In the second example, the proposed methodology is applied to update the stiffness parameters of the IASC-ASCE Phase II Simulated SHM Benchmark structure. The benchmark model is a four-story, two-bay by two-bay steel braced-frame finite-element structural model with 216 DOFs. A diagram for this model is shown in Figure 4 along with its dimensions, in which *x*-direction is the strong direction of the columns. A description of the benchmark problem including detailed nominal properties of the structural elements in the analytical model can be found in [16, 17, 27].

For steel-frame structures, earthquake structural damage usually occurs in braces and beam-column connections due to bucking and fracture, respectively. The case studies in the Phase II Simulated SHM Benchmark problem cover detection and assessment of both damage types but in this work, we focus on the brace damage cases. For brace damage, four damage cases are considered that are simulated by reducing the Young's moduli of certain braces in the structural model: 1) DP1B: 50% stiffness reduction in br 1-11 and br 7-17; 2) DP2B: like DP1B but 25% stiffness reduction in br1-11 and br7-17; 3) DP3B: same as DP1B, but in addition 25% stiffness reduction in br 19-29 and br 25-35; 4) DP3Bu: 50% and 25% stiffness reduction in br 1-11 and br 19-29. Here, br X-Y denotes the brace joining the nodes X and Y in Figure 4. The dashed lines in Figure 5 also indicate the corresponding damaged braces.

Both the full-sensor and partial-sensor scenarios are considered in the experiment. For the full-sensor scenario, measurements are available at the center of each side at each floor with the directions parallel to the side in either the positive $x$ direction or $y$ direction. For the partial-sensor scenario, only the measurements at the third floor (mid-height) and the roof are available.

To generate the simulated test data, a Matlab program for the simulated Phase II Benchmark [16, 17] is utilized and measurement noise equal to 10% RMS of the signal at the measured DOFs are added. For the monitoring stage, ten time-history segments ($q = 10$) of duration 20s (10,000 sampling points with sampling frequencies of 500 Hz) are generated to yield ten sets of independent MAP estimates of the experimental modal parameters ($\hat{\boldsymbol{\omega}}_r^2$ and $\hat{\boldsymbol{\psi}}_r$, $r = 1, \ldots, 10$) for all damage cases. Ching [16] identified a total of eight modes, four in the strong ($x$) direction and four in the weak ($y$) direction, of the structure by applying the MODE-ID methodology [15, 28] to each segment of the time histories. The identified MAP modal parameters are presented in [16, 17]. For the calibration stage, in order to get more accurate inferred MAP estimates of stiffness parameters, we increase the number of time-history segments for tests of the undamaged structure to one hundred ($q = 100$), and generate the 200s time-history by changing the “time_duration” to 200s in the Matlab program for the simulated Phase II Benchmark to identify one hundred sets of independent MAP estimates of the modal parameters. In practice when performing modal identification, some lower modes might not be detected and the order of the modes might switch when damage occurs, but this is not a problem for the proposed method because it does not require matching the experimental and structural model modes.

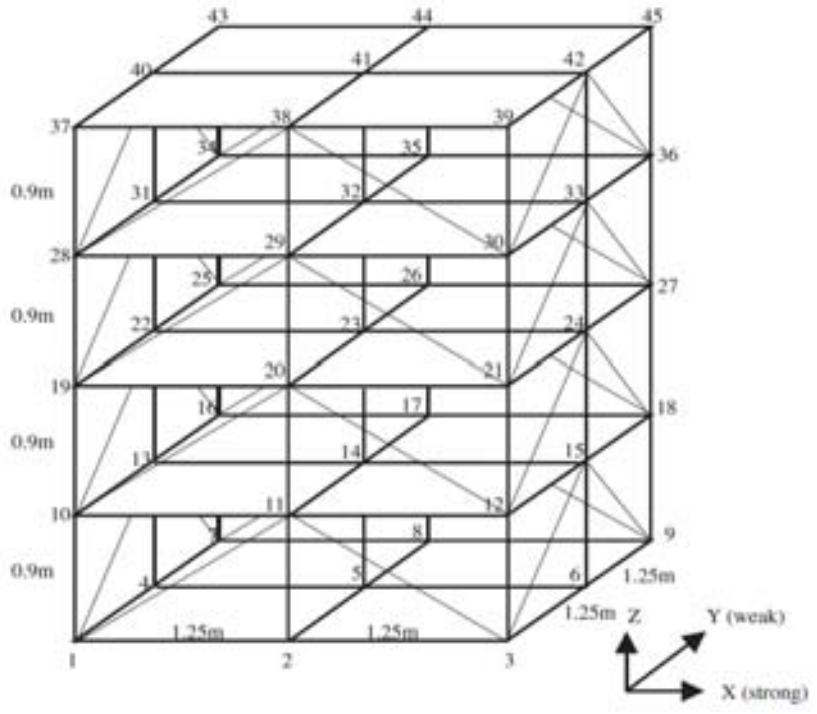

**Fig. 4.** The diagram of the benchmark structure [16,17].

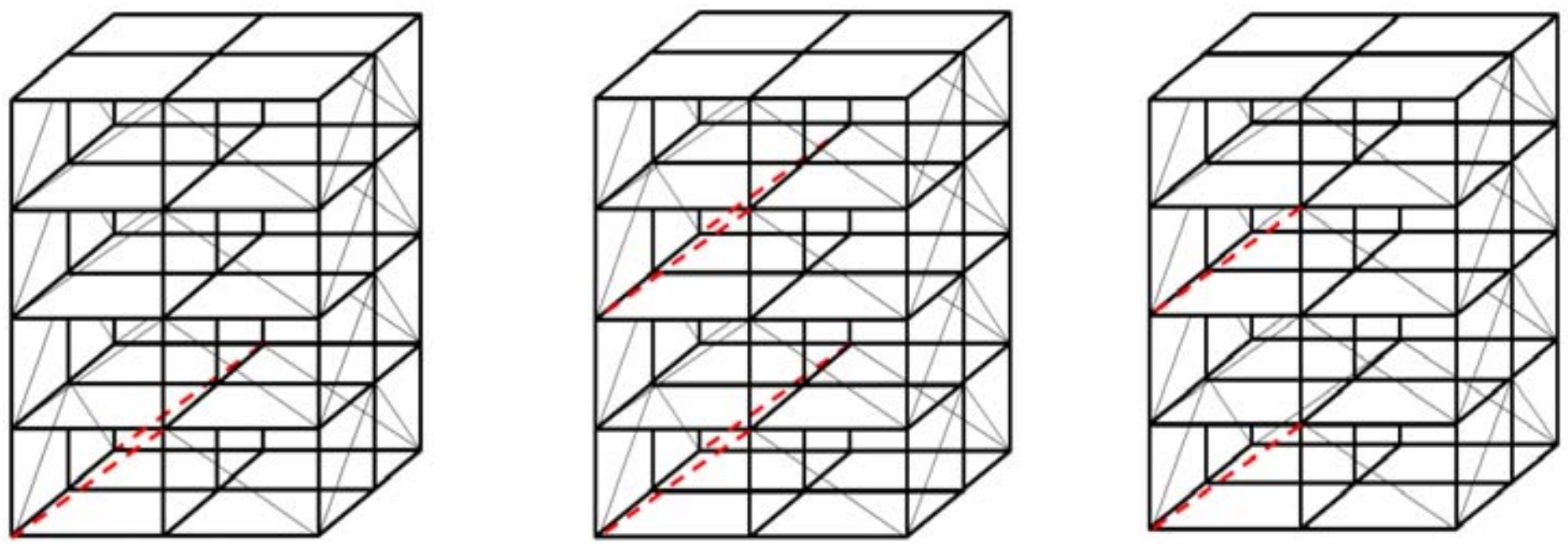

(a) DP1B and DP2B (b) DP3B (c) DP3BU

**Fig. 5.** Damage patterns for brace-damage cases (the dashed lines indicate the corresponding damaged locations) [16,17].

*4.2.2. Damage detection and assessment*

For brace damage cases, a 3-D 12-DOF shear-building model with rigid floors and three DOF per floor (translations parallel to the $x$-and $y$-axes and rotation about the $z$-axis) is employed in the damage inference procedure. The mass matrix is obtained by projecting the full mass matrix for the 216 DOF finite element structural model onto the 12 DOF and is taken as known during the system identification procedure. To locate the faces sustaining brace damage, the stiffness matrix $\mathbf{K}$ is parameterized as

$$\mathbf{K}(\boldsymbol{\theta}) = \mathbf{K}_0 + \sum_u \sum_v \theta_{uv} \bar{\mathbf{K}}_{uv} \tag{55}$$

where $u = 1, \ldots, 4$ refers to the story number and $v = +x, -x, +y, -y$ indicates the direction of the outward normal of each face at each floor. The $\bar{\mathbf{K}}_{uv}'s$ are the 'nominal' stiffness matrices computed based on shear-building assumptions. Four stiffness scaling parameters are used for each story to give a stiffness scaling parameter vector $\boldsymbol{\theta}$ with 16 components, corresponding to 16 substructures (four faces of four stories). Note that the relative stiffness loss for a particular face with 25% and 50% stiffness loss in any one brace is computed to be 5.7% and 11.3%, respectively. This corresponds to a stiffness scaling parameter of 94.3% and 88.7%, respectively, of the undamaged values.

During the calibration stage, we utilize Algorithm 1 based on the identified modal parameters from tests of the undamaged structure to find the MAP structural stiffness scaling parameters $\hat{\boldsymbol{\theta}}_u$ and their corresponding c.o.v, which are shown in Tables 6 and 7 for the two sensor scenarios. Similar with the results in Example 1, identification results for the MAP estimates of all stiffness scaling parameters $\theta_j$ are accurate, i.e, errors are smaller than 0.6% and 4.2% for full sensor and partial sensor scenarios, respectively, when the number of data segments is selected to be 100 ($q = 100$). During the monitoring stage, we choose the MAP value $\hat{\boldsymbol{\theta}}_u$ from the calibration stage as pseudo-data for $\boldsymbol{\theta}$ and run Algorithm 2 based on ten sets of identified modal parameters as the primary data. The stiffness ratios of the MAP estimates of the stiffness scaling parameters $\tilde{\theta}_{uv}$ with respect to those inferred from the calibration stage and their associated c.o.v. are also tabulated in Tables 6 and 7 for the two sensor scenarios.

In Tables 6 and 7, the actual damaged locations are made bold for comparison. It is observed that most of components with non-bold font have their stiffness ratio exactly equal to one, showing they are unchanged by the monitoring stage data, and the corresponding c.o.v. for each of these components is zero, which means these substructures have no stiffness reduction with full confidence (conditional on the modeling) compared with that of the calibration stage. If we issue a damage alarm for a substructure when the corresponding stiffness ratio is smaller than one (stiffness loss is larger than zero), it is seen that no false-negative or false-positive damage indications are produced for either of the sensor scenarios. There are two stiffness scaling parameters, $\theta_{2,+x}$ for DP3Bu.fs case and $\theta_{2,-x}$ for DP3Bu.ps case, with the stiffness ratio larger than one (1.002); however, since they show a very small *increase* in stiffness, they are not indicative of damage. Therefore, no threshold is required for damage identification in the proposed method, at least in the simulated data case. For the identified damage extent, we observe that the identified stiffness ratios are close to their actual values (0.943 and 0.887) for both the full sensor and partial sensor scenarios, and it is not surprising that the identified values are more accurate for the full sensor scenario.

**Table 6**

Stiffness ratios for the full-sensor scenario in simulated Phase II benchmark.

| Para-meter | RB.fs* | | DP1B.fs | | DP2B.fs | | DP3B.fs | | DP3Bu.fs | |
|---|---|---|---|---|---|---|---|---|---|---|
| | MAP value $(\hat{\boldsymbol{\theta}}_u)$ | c.o.v. (%) | MAP ratio $(\tilde{\boldsymbol{\theta}}/\hat{\boldsymbol{\theta}}_u)$ | c.o.v. (%) | MAP ratio $(\tilde{\boldsymbol{\theta}}/\hat{\boldsymbol{\theta}}_u)$ | c.o.v. (%) | MAP ratio $(\tilde{\boldsymbol{\theta}}/\hat{\boldsymbol{\theta}}_u)$ | c.o.v. (%) | MAP ratio $(\tilde{\boldsymbol{\theta}}/\hat{\boldsymbol{\theta}}_u)$ | c.o.v. (%) |
| $\theta_{1,+x}$ | 1.002 | 0.435 | 1.000 | 0.000 | 1.000 | 0.000 | 1.000 | 0.000 | 1.000 | 0.000 |
| $\theta_{2,+x}$ | 0.999 | 0.535 | 1.000 | 0.000 | 1.000 | 0.000 | 1.000 | 0.000 | 1.002 | 0.048 |
| $\theta_{3,+x}$ | 1.000 | 0.748 | 1.000 | 0.000 | 1.000 | 0.000 | 1.000 | 0.000 | 1.000 | 0.000 |
| $\theta_{4,+x}$ | 0.996 | 0.510 | 1.000 | 0.000 | 1.000 | 0.000 | 1.000 | 0.000 | 1.000 | 0.000 |
| $\theta_{1,+y}$ | 0.999 | 0.221 | **0.889** | **0.158** | **0.940** | **0.154** | **0.885** | **0.159** | 1.000 | 0.000 |
| $\theta_{2,+y}$ | 0.998 | 0.281 | 1.000 | 0.000 | 1.000 | 0.000 | 1.000 | 0.000 | 1.000 | 0.000 |
| $\theta_{3,+y}$ | 1.002 | 0.373 | 1.000 | 0.000 | 1.000 | 0.000 | **0.946** | **0.069** | 1.000 | 0.000 |
| $\theta_{4,+y}$ | 0.998 | 0.275 | 1.000 | 0.000 | 1.000 | 0.000 | 1.000 | 0.000 | 1.000 | 0.000 |
| $\theta_{1,-x}$ | 0.994 | 0.230 | 1.000 | 0.000 | 1.000 | 0.000 | 1.000 | 0.000 | 1.000 | 0.000 |
| $\theta_{2,-x}$ | 0.999 | 0.233 | 1.000 | 0.000 | 1.000 | 0.000 | 1.000 | 0.000 | 1.000 | 0.000 |
| $\theta_{3,-x}$ | 1.006 | 0.388 | 1.000 | 0.000 | 1.000 | 0.000 | 1.000 | 0.000 | 1.000 | 0.000 |
| $\theta_{4,-x}$ | 1.001 | 0.216 | 1.000 | 0.000 | 1.000 | 0.000 | 1.000 | 0.000 | 1.000 | 0.000 |
| $\theta_{1,-y}$ | 1.003 | 0.172 | **0.885** | **0.155** | **0.947** | **0.149** | **0.878** | **0.156** | **0.887** | **0.149** |
| $\theta_{2,-y}$ | 1.004 | 0.227 | 1.000 | 0.000 | 1.000 | 0.000 | 1.000 | 0.000 | 1.000 | 0.000 |
| $\theta_{3,-y}$ | 0.995 | 0.302 | 1.000 | 0.000 | 1.000 | 0.000 | **0.948** | **0.064** | **0.950** | **0.058** |
| $\theta_{4,-y}$ | 1.003 | 0.226 | 1.000 | 0.000 | 1.000 | 0.000 | 1.000 | 0.000 | 1.000 | 0.000 |

*.fs denotes the full-sensor scenario.

**Table 7**

Stiffness ratios for the partial-sensor scenario in simulated Phase II benchmark.

| Para-meter | RB.ps* | | DP1B.ps | | DP2B.ps | | DP3B.ps | | DP3Bu.ps | |
|---|---|---|---|---|---|---|---|---|---|---|
| | MAP value $(\hat{\boldsymbol{\theta}}_u)$ | c.o.v. (%) | MAP ratio $(\tilde{\boldsymbol{\theta}}/\hat{\boldsymbol{\theta}}_u)$ | c.o.v. (%) | MAP ratio $(\tilde{\boldsymbol{\theta}}/\hat{\boldsymbol{\theta}}_u)$ | c.o.v. (%) | MAP ratio $(\tilde{\boldsymbol{\theta}}/\hat{\boldsymbol{\theta}}_u)$ | c.o.v. (%) | MAP ratio $(\tilde{\boldsymbol{\theta}}/\hat{\boldsymbol{\theta}}_u)$ | c.o.v. (%) |
| $\theta_{1,+x}$ | 1.047 | 0.406 | 1.000 | 0.000 | 1.000 | 0.000 | 1.000 | 0.000 | 1.000 | 0.000 |
| $\theta_{2,+x}$ | 0.997 | 0.534 | 1.000 | 0.000 | 1.000 | 0.000 | 1.000 | 0.000 | 1.000 | 0.000 |
| $\theta_{3,+x}$ | 0.958 | 0.808 | 1.000 | 0.000 | 1.000 | 0.000 | 1.000 | 0.000 | 1.000 | 0.000 |
| $\theta_{4,+x}$ | 0.996 | 0.505 | 1.000 | 0.000 | 1.000 | 0.000 | 1.002 | 0.047 | 1.000 | 0.000 |
| $\theta_{1,+y}$ | 1.007 | 0.220 | **0.909** | **0.162** | **0.946** | **0.149** | **0.870** | **0.164** | 1.000 | 0.000 |
| $\theta_{2,+y}$ | 0.997 | 0.282 | 1.000 | 0.000 | 1.000 | 0.000 | 1.000 | 0.000 | 1.000 | 0.000 |
| $\theta_{3,+y}$ | 0.996 | 0.374 | 1.000 | 0.000 | 1.000 | 0.000 | **0.946** | **0.070** | 1.000 | 0.000 |
| $\theta_{4,+y}$ | 0.995 | 0.277 | 1.000 | 0.000 | 1.000 | 0.000 | 1.000 | 0.000 | 1.000 | 0.000 |
| $\theta_{1,-x}$ | 0.973 | 0.242 | 1.000 | 0.000 | 1.000 | 0.000 | 1.000 | 0.000 | 1.000 | 0.000 |
| $\theta_{2,-x}$ | 1.002 | 0.232 | 1.000 | 0.000 | 1.002 | 0.075 | 1.000 | 0.000 | 1.002 | 0.077 |
| $\theta_{3,-x}$ | 1.023 | 0.379 | 1.000 | 0.000 | 1.000 | 0.000 | 1.000 | 0.000 | 1.000 | 0.000 |
| $\theta_{4,-x}$ | 1.002 | 0.214 | 1.000 | 0.000 | 1.000 | 0.000 | 1.000 | 0.000 | 1.000 | 0.000 |
| $\theta_{1,-y}$ | 1.016 | 0.169 | **0.939** | **0.156** | **0.941** | **0.148** | **0.890** | **0.160** | **0.885** | **0.145** |
| $\theta_{2,-y}$ | 1.004 | 0.226 | 1.000 | 0.000 | 1.000 | 0.000 | 1.000 | 0.000 | 1.000 | 0.000 |
| $\theta_{3,-y}$ | 0.982 | 0.306 | 1.000 | 0.000 | 1.000 | 0.000 | **0.949** | **0.065** | **0.951** | **0.057** |
| $\theta_{4,-y}$ | 1.002 | 0.225 | 1.000 | 0.000 | 1.000 | 0.000 | 1.000 | 0.000 | 1.000 | 0.000 |

*.ps denotes the partial-sensor scenario.

In Figure 6, the proposed Algorithm 1 and Algorithm 2 are compared with the method by Yuen et al. [18] by presenting the stiffness reduction ratios for various damage cases, where these ratios are defined as the difference between each MAP value $\hat{\theta}_{uv}$ of the stiffness scaling parameter from the calibration stage and the MAP value $\tilde{\theta}_{uv}$ from the current monitoring stage, normalized by $\hat{\theta}_{uv}$. For the method by Yuen et al. [18], the required hyper-parameters are not optimized based on the data but are, instead, determined a priori using judgment. By comparing the results in Figure 6, it is seen that Algorithm 1 gives more accurate stiffness reduction ratios than the method by Yuen et al. [18], especially in the partial-sensor case. This shows the benefit of the hierarchical Bayesian modeling and learning of the hyper-priors. The results for Algorithm 2, which further optimizes the hyper-parameters $\boldsymbol{\alpha}$, $\lambda$ and $\zeta$, show the substantial improvement in accuracy that comes from exploiting damage sparseness; there are no false or missed damage indications. Algorithm 2 therefore has the ability to allow much higher-resolution damage localization than the method by Yuen et al. [18]. Thus, it is concluded that the proposed hierarchical sparse Bayesian learning framework is a very effective strategy for structural health monitoring.

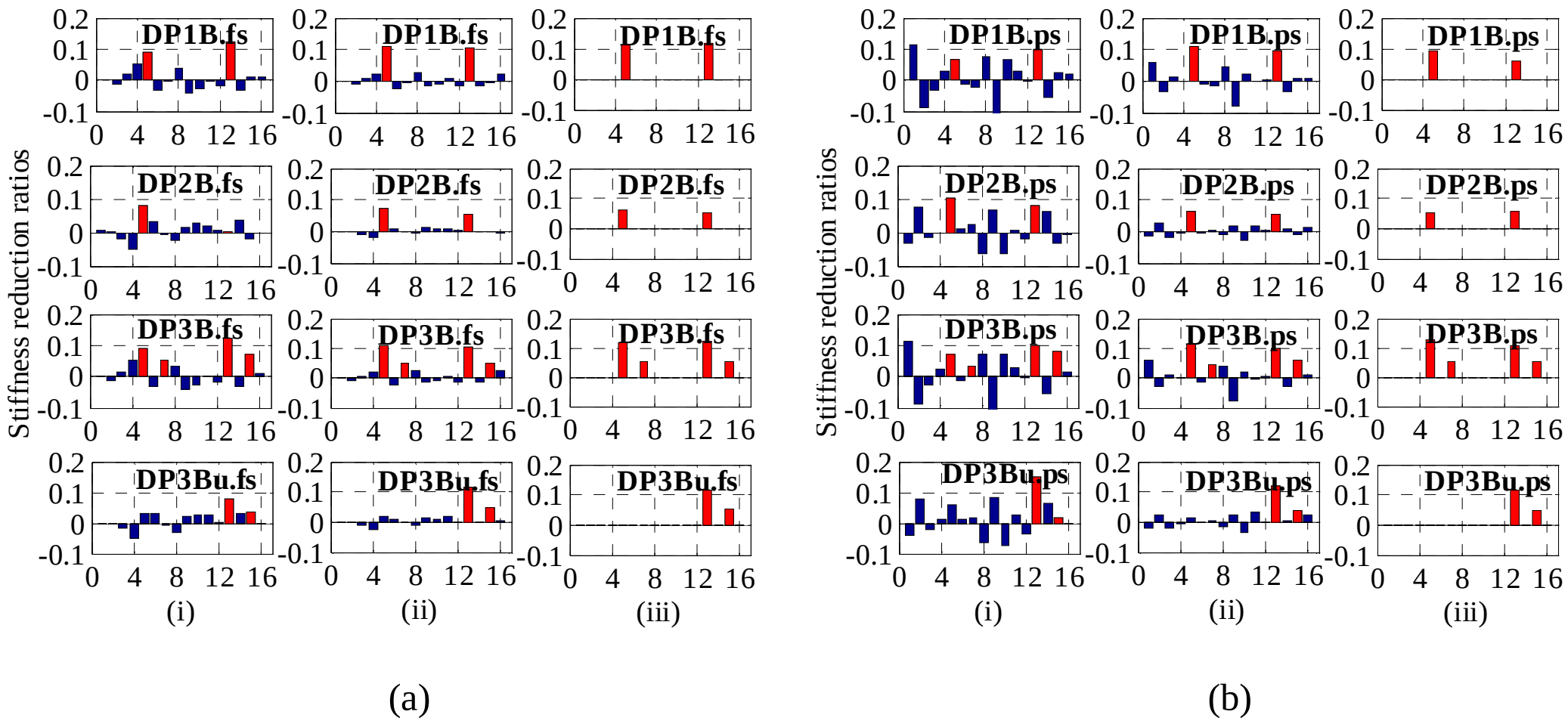

**Fig. 6.** Comparison of stiffness reduction ratios of three methods for the 16 substructures: (i) The method by Yuen et al. [18]; (ii) The proposed hierarchical method without utilizing damage sparseness (Algorithm 1); (iii) The proposed method utilizing sparseness (Algorithm 2); using the IASC-ASCE Phase II Simulated Benchmark data for four damage scenarios and two instrumentation scenarios: (a) full-sensor; (b) partial-sensor (red bars indicate actual damage locations).

To demonstrate the effect of hyper-prior $p(\boldsymbol{\alpha}|\lambda)$ in (12) on the spatial sparseness in the inferred stiffness reductions, the results of stiffness reduction ratios from running Algorithm 2 with different choices of hyper-parameter $\lambda$ are given in Figures 7(a)(i)-(d)(i) for the partial-sensor scenario. Note that $\lambda = 0$ corresponds to the original sparse Bayesian learning formulation with a uniform hyper-prior over $\boldsymbol{\alpha}$ as in Tipping (2001). By increasing the value of $\lambda$, fewer components or smaller values of $\boldsymbol{\theta}$ show a stiffness reduction. When $\lambda$ is sufficiently large, the inferred damage pattern becomes overly sparse, as shown by the third and fourth rows in Figure 7 (a)(i)-(d)(i),which also givens the value of $\lambda$ for which the change in values of the stiffness reduction ratios occurs. Recall that the actual damage can be accurately identified with no false-positive and false-negative alarms using Algorithm 2, which optimizes all hyper-parameters, and gives the results shown in the final rows of Figure 7 (a)(i)-(d)(i), including the optimal value of $\lambda$, that were previously shown in Figure 6. Therefore, the automatic estimation of $\lambda$ in Algorithm 2 is reliable for inferring accurate stiffness reductions with an appropriate sparseness level. This is a very useful advantage of the proposed algorithm since no user intervention is required to select $\lambda$.

In addition, we consider another case by defining a hyper-prior $p(\boldsymbol{\gamma}|\kappa)$ over the precision parameters, $\gamma_j = \alpha_j^{-1}, j = 1, \dots, n$, as usually used in sparse Bayesian learning [19]. The corresponding optimization of hyper-parameters $\boldsymbol{\gamma}$ and $\kappa$, as derived from the evidence maximization strategy, is in Appendix B. We run Algorithm 2 incorporating the alternative hyper-prior and its optimization and present the results in Figures 7(a)(ii)-(d)(ii). It is found that many components of $\boldsymbol{\theta}$ that correspond to undamaged substructures are changed during the updating, especially when the value of $\kappa$ is large. This hyper-prior on the precision parameters is therefore not as effective as the choice of hyper-prior on the variance $\boldsymbol{\alpha}$ defined in (12).

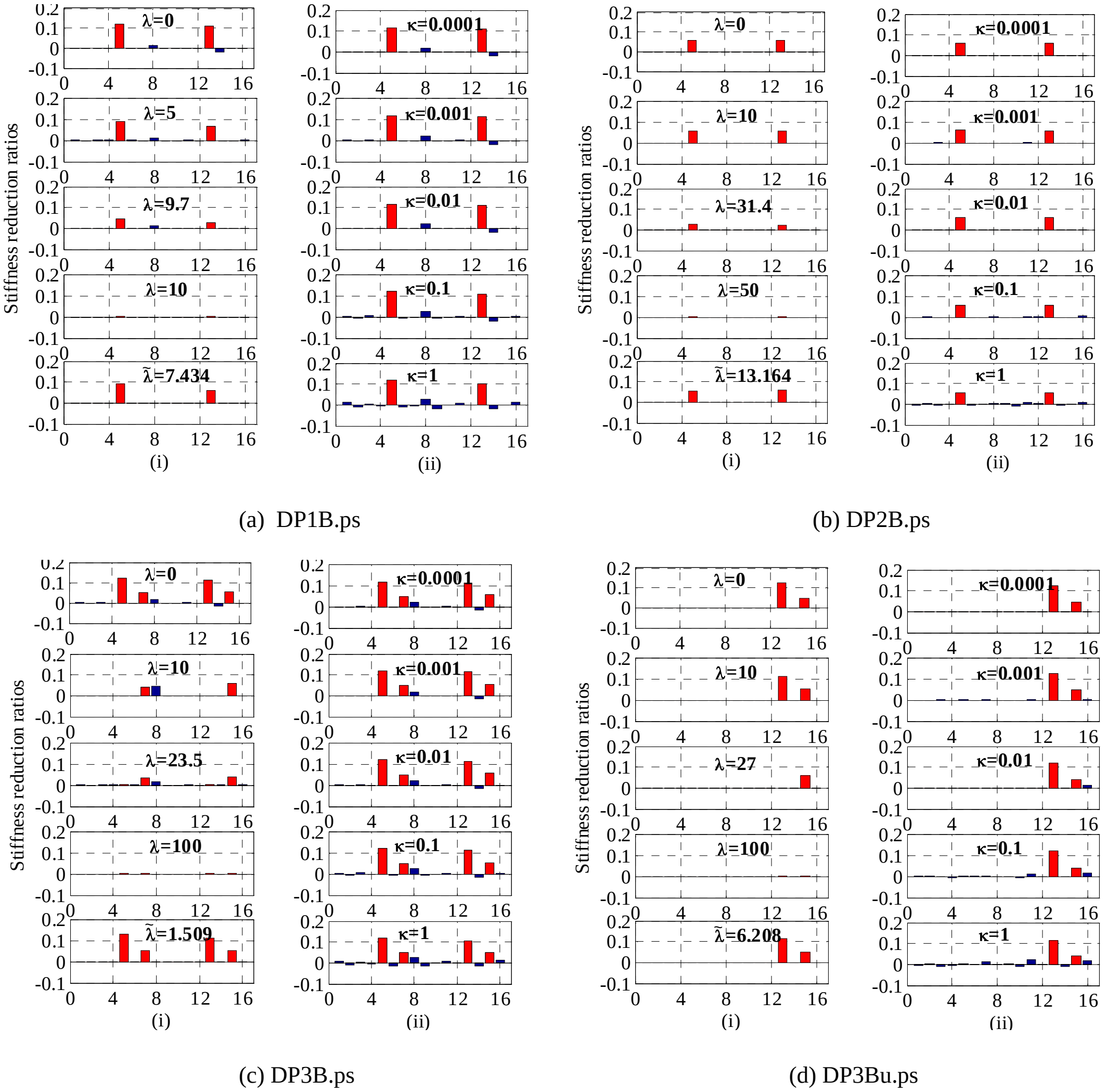

**Fig. 7.** Comparison of stiffness reduction ratios for the 16 substructures using two different hyper-priors: (i) $p(\boldsymbol{\alpha}|\lambda)$ defined in (12) and (ii) $p(\boldsymbol{\gamma}|\kappa)$ in (B2), using the IASC-ASCE Phase II Simulated Benchmark partial-sensor data, and for the damage patterns: (a) DP1B.ps; (b) DP2B.ps; (c) DP3B.ps and (d) DP3Bu.ps (red bars indicate actual damage locations).

To further portray the damage, the probability of damage for all sub-structures for different severities is calculated using Eq. (51). The damage probability curves for the sixteen stiffness scaling parameters $\theta_{uv}$ are shown in Figure 8 for the partial-sensor scenario case. All the actual damaged substructures are clearly shown to have a damage probability of almost unity with a large damage extent. Consider the $-y$ face of the first story as an example. For cases DP3B.ps and DP3Bu.ps, this substructure has a damage probability of almost unity for a stiffness loss ratio less than 10.0% and a probability of almost zero that the damage exceeds 12% (its actual value is 11.2%). However, for the case DP1B.ps, this substructure is inferred to have a damage probability of almost unity for a stiffness loss ratio less than 6.0% and a probability of almost zero that the damage exceeds 7%, showing an under-estimation of the damage extent. For the case of DP2B.ps, this substructure has a damage probability of almost unity for a damage extent of less than 5% and a probability of almost zero when the stiffness loss ratio exceeds 6.2% (its actual value is 5.7%). For the undamaged substructures, probabilities become zero when the damage exceeds 0.5%, showing a very small plausibility that damage has happened in these substructures. Compared with existing Bayesian updating methods [2,3,12,16-18], the confidence for correct damage indication is very high in the proposed method. The reduced uncertainty for the inferred stiffness scaling parameter **θ** from the proposed hierarchical sparse Bayesian learning framework helps suppress the occurrence of false and missed damage indications and increases the confidence of correct damage indications.

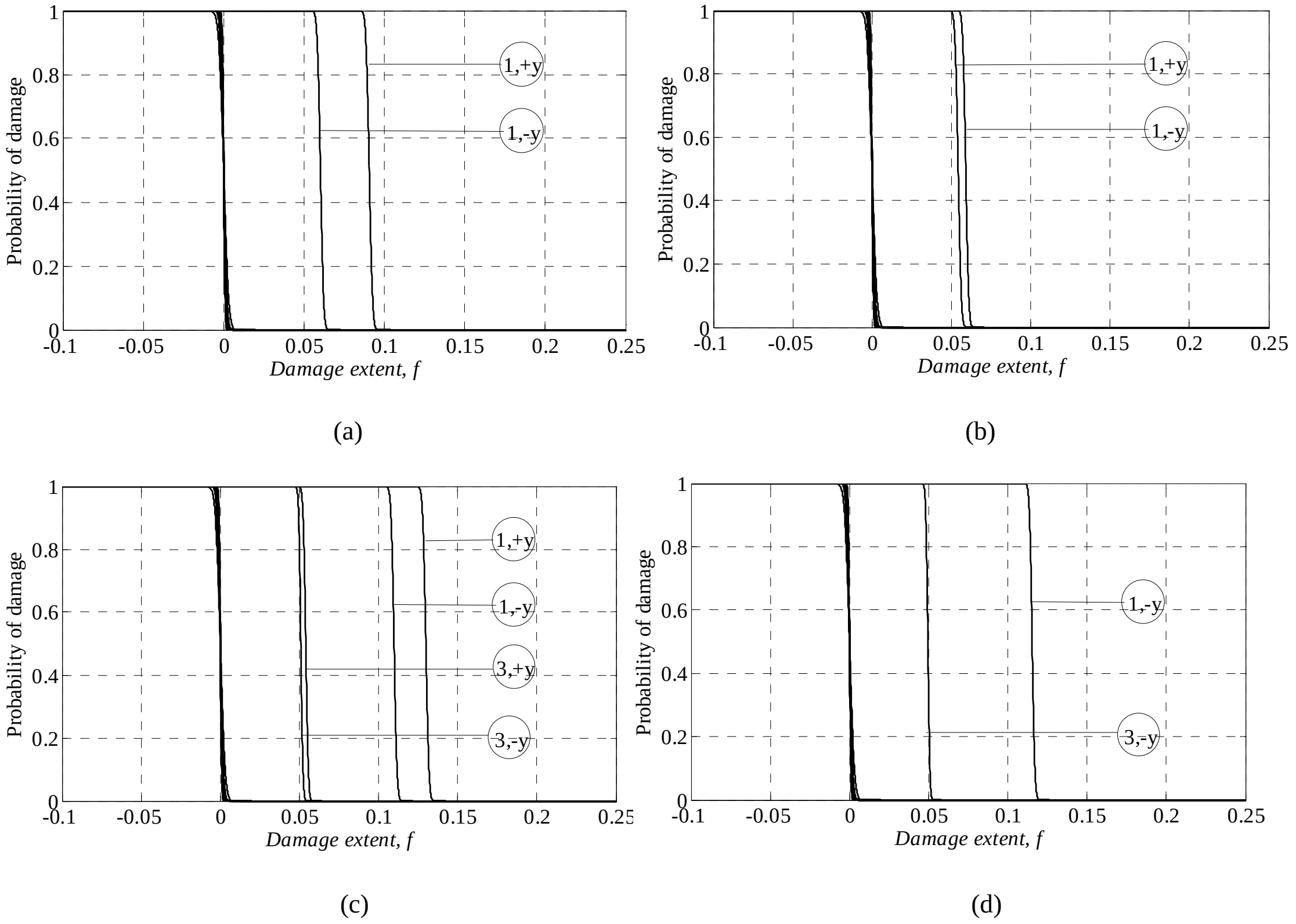

**Fig. 8.** Probability of damage exceeding *f* for the 16 substructures using the IASC-ASCE Phase II Simulated Benchmark data: (a) DP1B.ps; (b) DP2B.ps; (c) DP3B.ps and (d) DP3Bu.ps.

## 5. Conclusions

A new hierarchical sparse Bayesian learning methodology for probabilistic structural health monitoring with noisy incomplete modal data has been proposed. The method employs system modal parameters of the structure as extra variables and a multi-level hierarchical Bayesian model is constructed. Rather than directly solving the challenging nonlinear inverse problem related to eigenvalue equation, the proposed formulation applies an efficient iterative procedure that involves a series of coupled linear regression problems and provides a tractable form for the sparse Bayesian learning. The new algorithm estimates all uncertain hyper-parameters solely from the data, giving an algorithm for which no user-intervention is needed. The illustrative examples confirm the effectiveness and robustness of the new approach.

For the first example, comparison of the results with those of the state-of-the-art Bayesian updating method in [18], demonstrates the ability of the proposed method to update a structural model during a calibration stage for an undamaged building, showing the benefit of the hierarchical Bayesian modeling and learning of the hyper-parameters. The second example using the IASC-ASCE Phase II Simulated Benchmark data shows that for all cases, the simulated damage under study is reliably detected and the accuracy of the identified stiffness reduction is greatly enhanced by exploiting damage sparseness. Compared with the Bayesian updating method in [18], the occurrence of false-positive and false-negative damage detection in the presence of modeling errors is more effectively suppressed by the proposed method. Both methods also have an important advantage for actual applications which is that they can update the structural stiffness efficiently based on the information in the modal data from dynamic testing without knowing if any significant modes are missing in the modal data set, or whether the ordering of the modes switches due to damage.

In future work, we would like to apply the proposed method to real data from a structure that was instrumented before and after damage occurred.

## Acknowledgements

This work was supported by the U.S. National Science Foundation under award number EAR-0941374 to the California Institute of Technology. This support is gratefully acknowledged by both authors. This research is also supported by grant from the National Natural Science Foundation of China (NSFC grant no. 51308161), which partially supported the first author and this support is gratefully acknowledged.

**Nomenclature**

$m$ = Number of extracted modes in the modal identification

$s$ = Number of measured degrees of freedom

$d$ = Number of degrees of freedom of the identification model

$q$ = Number of time segments of measured modal data

$n$ = Number of substructures considered

$\mathbf{M}, \mathbf{K}$ = Mass and stiffness matrices of structural model

$\boldsymbol{\theta}$ = Structural stiffness scaling parameter vector

$\boldsymbol{\phi}_i, \omega_i^2$ = System mode shape vector and system natural frequency of the $i^{th}$ mode

$\beta$ = Equation-error precision parameter

$\hat{\omega}_{r,i}^2, \hat{\boldsymbol{\psi}}_{r,i}$ = MAP estimates of system natural frequency and mode shape vector of the $i^{th}$ mode from the $r^{th}$ data segment from modal identification

$\hat{\boldsymbol{\theta}}_u$ = MAP estimate of $\boldsymbol{\theta}$ determined from the calibration test data

$\boldsymbol{\Gamma}$ = Matrix that picks the measured degrees of freedom from the system mode shape

$\mathbf{T}$ = Transformation matrix between the vectors of $q$ sets of identified natural frequencies $\hat{\boldsymbol{\omega}}^2$ and the system natural frequencies $\boldsymbol{\omega}^2$

$\boldsymbol{\alpha}$ = Prior variance parameter vector for structural stiffness scaling parameters $\boldsymbol{\theta}$

$\eta, \boldsymbol{\rho}$ = Measurement-error precision parameters for mode shapes and natural frequencies

$\nu, \tau_i, \lambda, \zeta$ =Rate parameters controlling the exponential prior distributions of $\eta, \rho_i, \alpha_j, \lambda$, respectively

$\boldsymbol{\delta} = [\boldsymbol{\alpha}^T, \lambda, \zeta]^T$

$\boldsymbol{\xi} = [\beta, (\boldsymbol{\omega}^2)^T, \boldsymbol{\rho}, \boldsymbol{\tau}, \boldsymbol{\phi}^T, \eta, \nu]^T$

**Appendix A. MAP estimation of the hyper-parameters α, λ and ζ using the evidence strategy**

The MAP estimates of $\boldsymbol{\alpha}$, $\lambda$ and $\zeta$ can be obtained by maximizing the logarithm function of the pseudo-evidence (45) without including the constants that do not depend on $\boldsymbol{\alpha}$, $\lambda$ and $\zeta$. The objective function is:

$$J(\boldsymbol{\alpha}, \lambda, \zeta) = -\frac{1}{2}\log\left|\mathbf{A} + \left(\tilde{\beta}\tilde{\mathbf{H}}^T\tilde{\mathbf{H}}\right)^{-1}\right| - \frac{1}{2}\left(\hat{\boldsymbol{\theta}}_u - \left(\tilde{\mathbf{H}}^T\tilde{\mathbf{H}}\right)^{-1}\tilde{\mathbf{H}}^T\tilde{\mathbf{b}}\right)^T \left(\mathbf{A} + \left(\tilde{\beta}\tilde{\mathbf{H}}^T\tilde{\mathbf{H}}\right)^{-1}\right)^{-1}\left(\hat{\boldsymbol{\theta}}_u - \left(\tilde{\mathbf{H}}^T\tilde{\mathbf{H}}\right)^{-1}\tilde{\mathbf{H}}^T\tilde{\mathbf{b}}\right)$$
$$+n\log\lambda - \lambda\sum_{j=1}^{n}\alpha_j + \log\zeta - \zeta\lambda \quad \text{(A1)}$$

where matrix $\tilde{\mathbf{H}}$, vector $\tilde{\mathbf{b}}$ and $\tilde{\beta}$ are to be evaluated at their MAP values.

$$\left|\mathbf{A}+\left(\tilde{\beta}\widetilde{\mathbf{H}}^T\widetilde{\mathbf{H}}\right)^{-1}\right| = \left|\left(\tilde{\beta}\widetilde{\mathbf{H}}^T\widetilde{\mathbf{H}}\right)^{-1}\right|\left|\tilde{\beta}\mathbf{A}\widetilde{\mathbf{H}}^T\widetilde{\mathbf{H}}+\mathbf{I}_n\right| = \left|\left(\tilde{\beta}\widetilde{\mathbf{H}}^T\widetilde{\mathbf{H}}\right)^{-1}\right||\mathbf{A}||\boldsymbol{\Sigma}_{\boldsymbol{\theta}}|^{-1} \tag{A2}$$

where $\boldsymbol{\Sigma}_{\boldsymbol{\theta}}$ is the covariance matrix of $\boldsymbol{\theta}$ shown after (46).

Then:

$$\log\left|\mathbf{A}+\left(\beta\widetilde{\mathbf{H}}^T\widetilde{\mathbf{H}}\right)^{-1}\right| = \log\left|\left(\tilde{\beta}\widetilde{\mathbf{H}}^T\widetilde{\mathbf{H}}\right)^{-1}\right| + \log|\mathbf{A}| + \log\left|\boldsymbol{\Sigma}_{\boldsymbol{\theta}}^{-1}\right| \tag{A3}$$

Using the Woodbury matrix identity, we get:

$$\left(\mathbf{A}+\left(\tilde{\beta}\widetilde{\mathbf{H}}^T\widetilde{\mathbf{H}}\right)^{-1}\right)^{-1} = \tilde{\beta}\widetilde{\mathbf{H}}^T\widetilde{\mathbf{H}} - \tilde{\beta}\widetilde{\mathbf{H}}^T\widetilde{\mathbf{H}}\left(\mathbf{A}^{-1}+\tilde{\beta}\widetilde{\mathbf{H}}^T\widetilde{\mathbf{H}}\right)^{-1}\tilde{\beta}\widetilde{\mathbf{H}}^T\widetilde{\mathbf{H}}$$

$$= \tilde{\beta}\widetilde{\mathbf{H}}^T\widetilde{\mathbf{H}} - \tilde{\beta}^2\widetilde{\mathbf{H}}^T\widetilde{\mathbf{H}}\boldsymbol{\Sigma}_{\boldsymbol{\theta}}\widetilde{\mathbf{H}}^T\widetilde{\mathbf{H}} \tag{A4}$$

Using $\boldsymbol{\Sigma}_{\boldsymbol{\theta}}^{-1} = \mathbf{A}^{-1}+\tilde{\beta}\widetilde{\mathbf{H}}^T\widetilde{\mathbf{H}}$ and $\widetilde{\boldsymbol{\theta}} = \boldsymbol{\Sigma}_{\boldsymbol{\theta}}\left(\tilde{\beta}\widetilde{\mathbf{H}}^T\tilde{\mathbf{b}}+\mathbf{A}^{-1}\widehat{\boldsymbol{\theta}}_u\right)$, we get:

$$\left(\widehat{\boldsymbol{\theta}}_u-\left(\widetilde{\mathbf{H}}^T\widetilde{\mathbf{H}}\right)^{-1}\widetilde{\mathbf{H}}^T\tilde{\mathbf{b}}\right)^T\left(\mathbf{A}+\left(\tilde{\beta}\widetilde{\mathbf{H}}^T\widetilde{\mathbf{H}}\right)^{-1}\right)^{-1}\left(\widehat{\boldsymbol{\theta}}_u-\left(\widetilde{\mathbf{H}}^T\widetilde{\mathbf{H}}\right)^{-1}\widetilde{\mathbf{H}}^{\mathbf{T}}\tilde{\mathbf{b}}\right)$$

$$= \left(\widehat{\boldsymbol{\theta}}_u-\left(\widetilde{\mathbf{H}}^T\widetilde{\mathbf{H}}\right)^{-1}\widetilde{\mathbf{H}}^T\tilde{\mathbf{b}}\right)^T\left(\tilde{\beta}\widetilde{\mathbf{H}}^T\widetilde{\mathbf{H}}-\tilde{\beta}^2\widetilde{\mathbf{H}}^T\widetilde{\mathbf{H}}\boldsymbol{\Sigma}_{\boldsymbol{\theta}}\widetilde{\mathbf{H}}^T\widetilde{\mathbf{H}}\right)\left(\widehat{\boldsymbol{\theta}}_u-\left(\widetilde{\mathbf{H}}^T\widetilde{\mathbf{H}}\right)^{-1}\widetilde{\mathbf{H}}^T\tilde{\mathbf{b}}\right)$$

$$= \tilde{\beta}\left(\widehat{\boldsymbol{\theta}}_{\mathbf{u}}-\left(\widetilde{\mathbf{H}}^{\mathbf{T}}\widetilde{\mathbf{H}}\right)^{-\mathbf{1}}\widetilde{\mathbf{H}}^{\mathbf{T}}\tilde{\mathbf{b}}\right)^{\mathbf{T}}\widetilde{\mathbf{H}}^{\mathbf{T}}\widetilde{\mathbf{H}}\left(\widehat{\boldsymbol{\theta}}_{\mathbf{u}}-\left(\widetilde{\mathbf{H}}^{\mathbf{T}}\widetilde{\mathbf{H}}\right)^{-\mathbf{1}}\widetilde{\mathbf{H}}^{\mathbf{T}}\tilde{\mathbf{b}}\right)$$

$$-\tilde{\beta}^2\left(\mathbf{H}\widehat{\boldsymbol{\theta}}_u-\mathbf{H}\left(\widetilde{\mathbf{H}}^T\widetilde{\mathbf{H}}\right)^{-1}\widetilde{\mathbf{H}}^T\tilde{\mathbf{b}}\right)^T\widetilde{\mathbf{H}}\boldsymbol{\Sigma}_{\boldsymbol{\theta}}\widetilde{\mathbf{H}}^T\left(\mathbf{H}\widehat{\boldsymbol{\theta}}_u-\mathbf{H}\left(\widetilde{\mathbf{H}}^T\widetilde{\mathbf{H}}\right)^{-1}\widetilde{\mathbf{H}}^T\tilde{\mathbf{b}}\right)$$

$$= \tilde{\beta}\left(\widehat{\boldsymbol{\theta}}_u-\left(\widetilde{\mathbf{H}}^T\widetilde{\mathbf{H}}\right)^{-1}\widetilde{\mathbf{H}}^T\tilde{\mathbf{b}}\right)^T\widetilde{\mathbf{H}}^T\widetilde{\mathbf{H}}\left(\widehat{\boldsymbol{\theta}}_u-\left(\widetilde{\mathbf{H}}^T\widetilde{\mathbf{H}}\right)^{-1}\widetilde{\mathbf{H}}^T\tilde{\mathbf{b}}-\tilde{\beta}\boldsymbol{\Sigma}_{\boldsymbol{\theta}}\left(\widetilde{\mathbf{H}}^T\widetilde{\mathbf{H}}\widehat{\boldsymbol{\theta}}_u-\widetilde{\mathbf{H}}^T\tilde{\mathbf{b}}\right)\right)$$

$$= \tilde{\beta}\left(\widehat{\boldsymbol{\theta}}_u-\left(\widetilde{\mathbf{H}}^T\widetilde{\mathbf{H}}\right)^{-1}\widetilde{\mathbf{H}}^T\tilde{\mathbf{b}}\right)^T\widetilde{\mathbf{H}}^T\widetilde{\mathbf{H}}\left(\widehat{\boldsymbol{\theta}}_u-\left(\widetilde{\mathbf{H}}^T\widetilde{\mathbf{H}}\right)^{-1}\widetilde{\mathbf{H}}^T\tilde{\mathbf{b}}-\boldsymbol{\Sigma}_{\boldsymbol{\theta}}\left(\tilde{\beta}\widetilde{\mathbf{H}}^T\widetilde{\mathbf{H}}\widehat{\boldsymbol{\theta}}_{\mathbf{u}}+\mathbf{A}^{-1}\widehat{\boldsymbol{\theta}}_u-\boldsymbol{\Sigma}_{\boldsymbol{\theta}}^{-1}\widetilde{\boldsymbol{\theta}}\right)\right)$$

$$= \tilde{\beta}\left(\widehat{\boldsymbol{\theta}}_u-\left(\widetilde{\mathbf{H}}^T\widetilde{\mathbf{H}}\right)^{-1}\widetilde{\mathbf{H}}^T\tilde{\mathbf{b}}\right)^{\mathbf{T}}\widetilde{\mathbf{H}}^T\widetilde{\mathbf{H}}\left(\widehat{\boldsymbol{\theta}}_u-\left(\widetilde{\mathbf{H}}^T\widetilde{\mathbf{H}}\right)^{-1}\widetilde{\mathbf{H}}^T\tilde{\mathbf{b}}-\left(\widehat{\boldsymbol{\theta}}_u-\widetilde{\boldsymbol{\theta}}\right)\right)$$

$$= \tilde{\beta}\left(\widehat{\boldsymbol{\theta}}_u-\left(\widetilde{\mathbf{H}}^T\widetilde{\mathbf{H}}\right)^{-1}\widetilde{\mathbf{H}}^T\tilde{\mathbf{b}}-\left(\widehat{\boldsymbol{\theta}}_u-\widetilde{\boldsymbol{\theta}}\right)\right)^T\widetilde{\mathbf{H}}^T\widetilde{\mathbf{H}}\left(\widehat{\boldsymbol{\theta}}_u-\left(\widetilde{\mathbf{H}}^T\widetilde{\mathbf{H}}\right)^{-1}\widetilde{\mathbf{H}}^T\tilde{\mathbf{b}}-\left(\widehat{\boldsymbol{\theta}}_u-\widetilde{\boldsymbol{\theta}}\right)\right)$$

$$+\tilde{\beta}\left(\widetilde{\mathbf{H}}^T\widetilde{\mathbf{H}}\widehat{\boldsymbol{\theta}}_u-\widetilde{\mathbf{H}}^T\tilde{\mathbf{b}}\right)^T\left(\widehat{\boldsymbol{\theta}}_u-\widetilde{\boldsymbol{\theta}}\right)-\tilde{\beta}\left(\widehat{\boldsymbol{\theta}}_u-\widetilde{\boldsymbol{\theta}}\right)^T\widetilde{\mathbf{H}}^T\widetilde{\mathbf{H}}\left(\widehat{\boldsymbol{\theta}}_u-\widetilde{\boldsymbol{\theta}}\right)$$

$$= \tilde{\beta}\left(\widehat{\boldsymbol{\theta}}_{\mathbf{u}}-\left(\widetilde{\mathbf{H}}^T\widetilde{\mathbf{H}}\right)^{-1}\widetilde{\mathbf{H}}^T\tilde{\mathbf{b}}-\left(\widehat{\boldsymbol{\theta}}_u-\widetilde{\boldsymbol{\theta}}\right)\right)^T\widetilde{\mathbf{H}}^T\widetilde{\mathbf{H}}\left(\widehat{\boldsymbol{\theta}}_u-\left(\widetilde{\mathbf{H}}^T\widetilde{\mathbf{H}}\right)^{-1}\widetilde{\mathbf{H}}^T\tilde{\mathbf{b}}-\left(\widehat{\boldsymbol{\theta}}_u-\widetilde{\boldsymbol{\theta}}\right)\right)$$

$$+\left(\tilde{\beta}\widetilde{\mathbf{H}}^{\mathbf{T}}\widetilde{\mathbf{H}}\widehat{\boldsymbol{\theta}}_u+\mathbf{A}^{-1}\widehat{\boldsymbol{\theta}}_{\mathbf{u}}-\boldsymbol{\Sigma}_{\boldsymbol{\theta}}^{-1}\widetilde{\boldsymbol{\theta}}\right)^T\left(\widehat{\boldsymbol{\theta}}_u-\widetilde{\boldsymbol{\theta}}\right)-\tilde{\beta}\left(\widehat{\boldsymbol{\theta}}_u-\widetilde{\boldsymbol{\theta}}\right)^T\widetilde{\mathbf{H}}^T\widetilde{\mathbf{H}}\left(\widehat{\boldsymbol{\theta}}_u-\widetilde{\boldsymbol{\theta}}\right)$$

$$= \tilde{\beta}\left(\widehat{\boldsymbol{\theta}}_u-\left(\widetilde{\mathbf{H}}^T\widetilde{\mathbf{H}}\right)^{-1}\widetilde{\mathbf{H}}^T\tilde{\mathbf{b}}-\left(\widehat{\boldsymbol{\theta}}_u-\widetilde{\boldsymbol{\theta}}\right)\right)^T\widetilde{\mathbf{H}}^T\widetilde{\mathbf{H}}\left(\widehat{\boldsymbol{\theta}}_u-\left(\widetilde{\mathbf{H}}^{\mathbf{T}}\widetilde{\mathbf{H}}\right)^{-1}\widetilde{\mathbf{H}}^{\mathbf{T}}\tilde{\mathbf{b}}-\left(\widehat{\boldsymbol{\theta}}_u-\widetilde{\boldsymbol{\theta}}\right)\right)$$

$$+\left(\widehat{\boldsymbol{\theta}}_u-\widetilde{\boldsymbol{\theta}}\right)^T\boldsymbol{\Sigma}_{\boldsymbol{\theta}}^{-1}\left(\widehat{\boldsymbol{\theta}}_u-\widetilde{\boldsymbol{\theta}}\right)-\beta\left(\widehat{\boldsymbol{\theta}}_u-\widetilde{\boldsymbol{\theta}}\right)^T\widetilde{\mathbf{H}}^T\widetilde{\mathbf{H}}\left(\widehat{\boldsymbol{\theta}}_u-\widetilde{\boldsymbol{\theta}}\right)$$

$$= \tilde{\beta}\left(\widetilde{\boldsymbol{\theta}}-\left(\widetilde{\mathbf{H}}^T\widetilde{\mathbf{H}}\right)^{-1}\widetilde{\mathbf{H}}^T\tilde{\mathbf{b}}\right)^T\widetilde{\mathbf{H}}^T\widetilde{\mathbf{H}}\left(\widetilde{\boldsymbol{\theta}}-\left(\widetilde{\mathbf{H}}^T\widetilde{\mathbf{H}}\right)^{-1}\widetilde{\mathbf{H}}^T\tilde{\mathbf{b}}\right)+\left(\widehat{\boldsymbol{\theta}}_u-\widetilde{\boldsymbol{\theta}}\right)^T\mathbf{A}^{-1}\left(\widehat{\boldsymbol{\theta}}_u-\widetilde{\boldsymbol{\theta}}\right) \tag{A5}$$

Then the derivative of the objective function in (A1) with respect to $\alpha_j$ is given by:

$$\frac{\partial J(\boldsymbol{\alpha},\lambda,\zeta)}{\partial \alpha_j}$$

$$= \frac{1}{2}\frac{\partial}{\partial \alpha_j}\left[-\log\left|\mathbf{A} + \left(\tilde{\beta}\tilde{\mathbf{H}}^T\tilde{\mathbf{H}}\right)^{-1}\right| - \left(\hat{\boldsymbol{\theta}}_u - \left(\tilde{\mathbf{H}}^T\tilde{\mathbf{H}}\right)^{-1}\tilde{\mathbf{H}}^T\tilde{\mathbf{b}}\right)^T\left(\mathbf{A} + \left(\tilde{\beta}\tilde{\mathbf{H}}^T\tilde{\mathbf{H}}\right)^{-1}\right)^{-1}\left(\hat{\boldsymbol{\theta}}_u - \left(\tilde{\mathbf{H}}^T\tilde{\mathbf{H}}\right)^{-1}\tilde{\mathbf{H}}^T\tilde{\mathbf{b}}\right) - 2\lambda\sum_{j=1}^{n}\alpha_j\right]$$

$$= \frac{1}{2}\left[-\alpha_j^{-1} + \alpha_j^{-2}(\boldsymbol{\Sigma}_{\boldsymbol{\theta}})_{jj} + \alpha_j^{-2}\left(\hat{\boldsymbol{\theta}}_u - \tilde{\boldsymbol{\theta}}\right)_j^2 - 2\lambda\right] \quad \text{(A6)}$$

Setting the derivative in (A6) to zero lead to the update formula given in (46).

Following the corresponding procedure for the parameter $\lambda$, we set the derivative of the objective function in (A1) with respect to $\lambda$ to be zero:

$$\frac{\partial J(\boldsymbol{\alpha},\lambda,\zeta)}{\partial \lambda} = \frac{n}{\lambda} - \sum_{i=1}^{n}\alpha_j - \zeta = 0 \quad \text{(A7)}$$

which gives the optimal estimate $\tilde{\lambda}$ as in (47).

Finally, the MAP estimate of $\zeta$ is obtained by setting the following derivative of the objective function:

$$\frac{\partial J(\boldsymbol{\alpha},\lambda,\zeta)}{\partial \zeta} = \frac{1}{\zeta} - \lambda \quad \text{(A8)}$$

which gives the MAP value $\tilde{\zeta}$ shown in (48).

**Appendix B. MAP estimation of the hyper-parameter α when using the exponential hyper-prior over each precision $\alpha_j^{-1}$**

Following Tipping (2001), a gamma hyper-prior is chosen for the precision parameters $\gamma_j = \alpha_j^{-1}$:

$$p(\gamma_j|\chi,\kappa) = \text{Gam}(\gamma_j|\chi,\kappa) = \frac{\kappa^\chi}{\Gamma(\chi)}\gamma_j{}^{\chi-1}exp(-\kappa\gamma_j) \quad \text{(B1)}$$

We fix $\chi = 1$ and so obtain the exponential hyper-prior for $\gamma_j$, which is the maximum entropy PDF with support $[0,\infty)$ for given mean $\kappa^{-1}$. Then the prior PDF over the precision vector $\boldsymbol{\gamma} = \left[\alpha_1^{-1},\ldots,\alpha_n^{-1}\right]$ becomes:

$$p(\boldsymbol{\gamma}|\kappa) = \kappa^n\exp\left(-\kappa\sum_{j=1}^{n}\gamma_j\right) = \kappa^n\exp\left(-\kappa\sum_{j=1}^{n}\alpha_j^{-1}\right) \quad \text{(B2)}$$

Following the corresponding procedure in Appendix A, we find the MAP value $\tilde{\boldsymbol{\alpha}}$ by maximizing:

$$p\left(\boldsymbol{\alpha}|\tilde{\boldsymbol{\xi}},\hat{\boldsymbol{\omega}}^2,\hat{\boldsymbol{\psi}},\tilde{\boldsymbol{\theta}}_u\right) \propto p\left(\hat{\boldsymbol{\theta}}_u|\tilde{\boldsymbol{\omega}}^2,\tilde{\boldsymbol{\phi}},\tilde{\beta},\boldsymbol{\alpha}\right)p(\boldsymbol{\gamma}|\kappa)$$

$$= \mathcal{N}\left(\hat{\boldsymbol{\theta}}_u\middle|\left(\tilde{\mathbf{H}}^T\tilde{\mathbf{H}}\right)^{-1}\tilde{\mathbf{H}}^T\tilde{\mathbf{b}},\mathbf{D}\right)\cdot\kappa^{-n}\exp\left(-\kappa\sum_{j=1}^{n}\alpha_j^{-1}\right) \quad \text{(B3)}$$

Maximizing the logarithm function of (B3) with respect to $\alpha_j$ gives the MAP estimate $\alpha_j$ as:

$$\tilde{\alpha}_j = (\boldsymbol{\Sigma}_{\boldsymbol{\theta}})_{jj} + (\hat{\boldsymbol{\theta}}_u - \tilde{\boldsymbol{\theta}})_j^2 + \kappa \quad \text{(B4)}$$